\documentclass[conference]{IEEEtran}
\usepackage{tikz}
\usepackage{amsmath}
\usepackage{listings}
\usepackage{xspace}
\usepackage{cite}

\usepackage{hyperref}
\usepackage{bigstrut}

\usepackage{filecontents}
\usepackage{booktabs}
\usepackage{amsthm}

\usepackage{savesym}

\usepackage{changes}

\usepackage{subcaption}
\usepackage{url}
\usepackage{amsfonts}
\usepackage[T1]{fontenc}
\usepackage{multirow}
\usepackage{algpseudocode}
\usepackage{graphicx}
\usepackage{enumitem}
\usepackage{numprint}
\usepackage[normalem]{ulem}
\usepackage{tcolorbox}
\usepackage{xurl}
\usepackage{color}

\definecolor{lightgray}{rgb}{.9,.9,.9}
\definecolor{darkgray}{rgb}{.4,.4,.4}
\definecolor{purple}{rgb}{0.65, 0.12, 0.82}

\lstdefinelanguage{JavaScript}{
  keywords={typeof, new, true, false, catch, function, return, null, catch, switch, var, if, in, while, do, else, case, break},
  keywordstyle=\color{blue}\bfseries,
  ndkeywords={class, export, boolean, throw, implements, import, this},
  ndkeywordstyle=\color{darkgray}\bfseries,
  identifierstyle=\color{black},
  sensitive=false,
  comment=[l]{//},
  morecomment=[s]{/*}{*/},
  commentstyle=\color{purple}\ttfamily,
  stringstyle=\color{red}\ttfamily,
  morestring=[b]',
  morestring=[b]"
}

\DeclareMathDelimiter{(}{\mathopen} {operators}{"28}{largesymbols}{"00}
\DeclareMathDelimiter{)}{\mathclose}{operators}{"29}{largesymbols}{"01}

\setitemize{noitemsep,topsep=0pt,parsep=0pt,partopsep=0pt}
\setdescription{noitemsep,topsep=0pt,parsep=0pt,partopsep=0pt}

\makeatletter
\g@addto@macro{\normalsize}{%
    \setlength{\abovedisplayskip}{5pt}
    \setlength{\abovedisplayshortskip}{5pt}
    \setlength{\belowdisplayskip}{5pt}
    \setlength{\belowdisplayshortskip}{5pt}}
\makeatother

\newcommand{\etal}{\textit{et al.\xspace}}
\npthousandsep{,}
\newcommand{\tool}{\textsc{DeFiPoser}\xspace}
\newcommand{\toolBF}{\textsc{DeFiPoser-ARB}\xspace} 
\newcommand{\toolZThree}{\textsc{DeFiPoser-SMT}\xspace}

\usepackage{tikz}

\newcommand{\circled}[1]{\raisebox{.5pt}{\textcircled{\raisebox{-.9pt} {{#1}}}}}

\newcommand{\point}[1]{\par\smallskip\noindent\textbf{#1:}\xspace}

\newcommand*\wrapletters[1]{\wr@pletters#1\@nil}
\def\wr@pletters#1#2\@nil{#1\allowbreak\if&#2&\else\wr@pletters#2\@nil\fi}

\hypersetup{
    colorlinks=true,
    linkcolor=blue,
    filecolor=magenta,
    urlcolor=blue,
    pdftitle={On the Just-In-Time Discovery of Profit-Generating Transactions in DeFi Protocols},
    pdfpagemode=FullScreen,
}

\usepackage{pdflscape}
\usepackage{rotating}
\usepackage{array}
\newcolumntype{L}{>{\centering\arraybackslash}m{3cm}}

\newcommand{\empirical}[1]{#1}

\newcommand{\powerset}{\raisebox{.15\baselineskip}{\Large\ensuremath{\wp}}}

\usepackage{numprint}
\usepackage{fp}
\npthousandsep{,}
\npdecimalsign{.}
\DeclareRobustCommand{\Ether}[1]{%
\FPeval{\ethamount}{round(#1, 2)}%
\FPeval{\ethprice}{400}%
\FPeval{\usdamount}{round(\ethamount * \ethprice, 0)}%
$\numprint{\ethamount}$~ETH ($\numprint{\usdamount}$~USD)\xspace}

\newcommand{\NumBlocks}{\empirical{$950,000$}\xspace}
\newcommand{\NumDays}{\empirical{$150$ days}\xspace}

\newcommand{\NumActions}{\empirical{$96$}\xspace}
\newcommand{\NumAssets}{\empirical{$25$}\xspace}
\newcommand{\NumERC}{\empirical{$24$}\xspace}
\newcommand{\StartingBlock}{\empirical{$9,100,000$}\xspace}
\newcommand{\EndingBlock}{\empirical{$10,050,000$}\xspace}
\newcommand{\FastGasPrice}{\empirical{$32$~GWei}\xspace}
\newcommand{\AverageBlockTime}{\empirical{$13.5$ seconds}\xspace}
\newcommand{\NumPaths}{\empirical{$600$}\xspace}
\newcommand{\NumConfirmedStrategiesZThree}{\empirical{$1,556$}\xspace}
\newcommand{\NumConfirmedStrategiesBF}{\empirical{$2,709$}\xspace}
\newcommand{\AverageAnalsysTimePerBlockZThree}{\empirical{$5.39$ seconds}\xspace}
\newcommand{\AverageAnalsysTimePerBlockBF}{\empirical{$6.43$ seconds}\xspace}
\newcommand{\CapitalRequrementWithoutFlashLoan}{\empirical{\Ether{150}}}
\newcommand{\CapitalRequrementWithFlashLoan}{\empirical{\Ether{0.4}}}

\newcommand{\MaxRevenueConfirmedZThree}{\empirical{\Ether{22.40}}}
\newcommand{\MaxRevenueConfirmedBF}{\empirical{\Ether{81.31}}}

\newcommand{\TotalRevenueUnconfirmedZThree}{\empirical{\Ether{3577.14}}}
\newcommand{\TotalRevenueConfirmedZThree}{\empirical{\Ether{1552.32}}}
\newcommand{\TotalRevenueConfirmedBF}{\empirical{\Ether{4103.22}}}

\newcommand{\WeeklyRevenueZThree}{\empirical{\Ether{72.44}}}
\newcommand{\WeeklyRevenueBF}{\empirical{\Ether{191.48}}}

\newcommand{\StateChangeBelowOneHundredPathsBlockPercentage}{\empirical{$32.71\%$}\xspace}
\newcommand{\ETHStateChange}{\empirical{$36.76\%$}\xspace}
\newcommand{\DAIStateChange}{\empirical{$14.62\%$}\xspace}
\newcommand{\POAStateChange}{\empirical{$0.08\%$}\xspace}
\newcommand{\ZFoundStrategies}{\empirical{$13,317$}\xspace}
\newcommand{\BFFoundStrategies}{\empirical{$2,709$}\xspace}

\newcommand{\BZXExceedsBlockReward}{\empirical{$874\times$}\xspace}
\newcommand{\ZTHREEExceedsBlockReward}{\empirical{$8.5\times$}\xspace}
\newcommand{\BFExceedsBlockReward}{\empirical{$31\times$}\xspace}

\usepackage[ruled,vlined]{algorithm2e}
\SetAlFnt{\small}

\usepackage{fancyvrb,newverbs,xcolor}
\definecolor{cverbbg}{gray}{0.93}

\newenvironment{lcverbatim}
 {\SaveVerbatim{cverb}}
 {\endSaveVerbatim
  \flushleft\fboxrule=0pt\fboxsep=.5em
  \colorbox{cverbbg}{%
    \makebox[\dimexpr\linewidth-2\fboxsep][l]{\BUseVerbatim{cverb}}%
  }
  \endflushleft
}

\newverbcommand{\cverb}
  {\setbox\verbbox\hbox\bgroup}
  {\egroup\colorbox{cverbbg}{\box\verbbox}}

\usepackage{todonotes}
\setuptodonotes{fancyline, inline, color=blue!30}

\usepackage[n,landau,notions,ff,mm]{cryptocode}
\definecolor{gamechangecolor}{gray}{0.74}

\definechangesauthor[name=Arthur, color=red]{arthur}

\definecolor{cblue}{rgb}{0.0, 0.28, 0.9}   
\definechangesauthor[name=Liyi, color=cblue]{liyi}

\begin{document}
\sloppy
\setcounter{tocdepth}{1}

\title{On the Just-In-Time Discovery of \\Profit-Generating Transactions in DeFi Protocols}

\author{
\IEEEauthorblockN{
Liyi Zhou, 
Kaihua Qin, 
Antoine Cully, 
Benjamin Livshits
and 
Arthur Gervais
}

\IEEEauthorblockA{
Imperial College London, United Kingdom\\
}
}


\maketitle

\begin{abstract}
Decentralized Finance~ (DeFi) is a blockchain-asset-enabled finance ecosystem with millions of daily USD transaction volume, billions of locked up USD, as well as a plethora of newly emerging protocols (for lending, staking, and exchanges). Because all transactions, user balances, and total value locked in DeFi are publicly readable, a natural question that arises is: how can we automatically craft profitable transactions across the intertwined DeFi platforms?

In this paper, we investigate two methods that allow us to automatically create profitable DeFi trades, one well-suited to arbitrage and the other applicable to more complicated settings. We first adopt the Bellman-Ford-Moore algorithm with \toolBF and then create logical DeFi protocol models for a theorem prover in \toolZThree. While \toolBF focuses on DeFi transactions that form a cycle and performs very well for arbitrage, \toolZThree can detect more complicated profitable transactions. We estimate that \toolBF and \toolZThree can generate an average weekly revenue of~\WeeklyRevenueBF and ~\WeeklyRevenueZThree respectively, with the highest transaction revenue being~\MaxRevenueConfirmedBF and~\MaxRevenueConfirmedZThree respectively. We further show that \toolZThree finds the known economic bZx attack from February~2020, which yields~$0.48$M USD. Our forensic investigations show that this opportunity existed for~$69$ days and could have yielded more revenue if exploited one day earlier. Our evaluation spans~\NumDays, given~\NumActions DeFi protocol actions, and~\NumAssets assets.

Looking beyond the financial gains mentioned above, \emph{forks} deteriorate the blockchain consensus security, as they increase the risks of double-spending and selfish mining. We explore the implications of \toolBF and \toolZThree on blockchain consensus. Specifically, we show that the trades identified by our tools exceed the Ethereum block reward by up to~\BZXExceedsBlockReward. Given optimal adversarial strategies provided by a Markov Decision Process~(MDP), we quantify the value threshold at which a profitable transaction qualifies as Miner Extractable Value~(MEV) and would incentivize MEV-aware miners to fork the blockchain. For instance, we find that on Ethereum, a miner with a hash rate of~$10\%$ would fork the blockchain if an MEV opportunity exceeds~$4\times$ the block reward.
\end{abstract}

\section{Introduction}
\label{sec:intro}

Blockchain-based decentralized finance protocols (commonly referred to as DeFi) have attracted a recent surge in popularity and value stored exceeding~$13$ billion USD. The currently most popular DeFi platforms are based on the Ethereum blockchain and its system of smart contracts, which regularly gives nascence to new applications, mirrored and inspired by the traditional centralized finance system. Examples are asset exchanges~\cite{dydx, uniswap2018}, margin trading~\cite{dydx, bzxnetwork}, lending/borrowing platforms~\cite{makerdao, compoundfinance}, and derivatives~\cite{compoundfinance}. DeFi, moreover, can surprise with novel use-cases such as constant product market maker exchanges~\cite{uniswap2018, balancer-finance} and flash loans~---~instant loans where the lender bears no risk that the borrower does not repay the loan~\cite{aave, dydx, qin2020attacking}.

\begin{figure}[tb]
\centering
\includegraphics[width = \columnwidth]{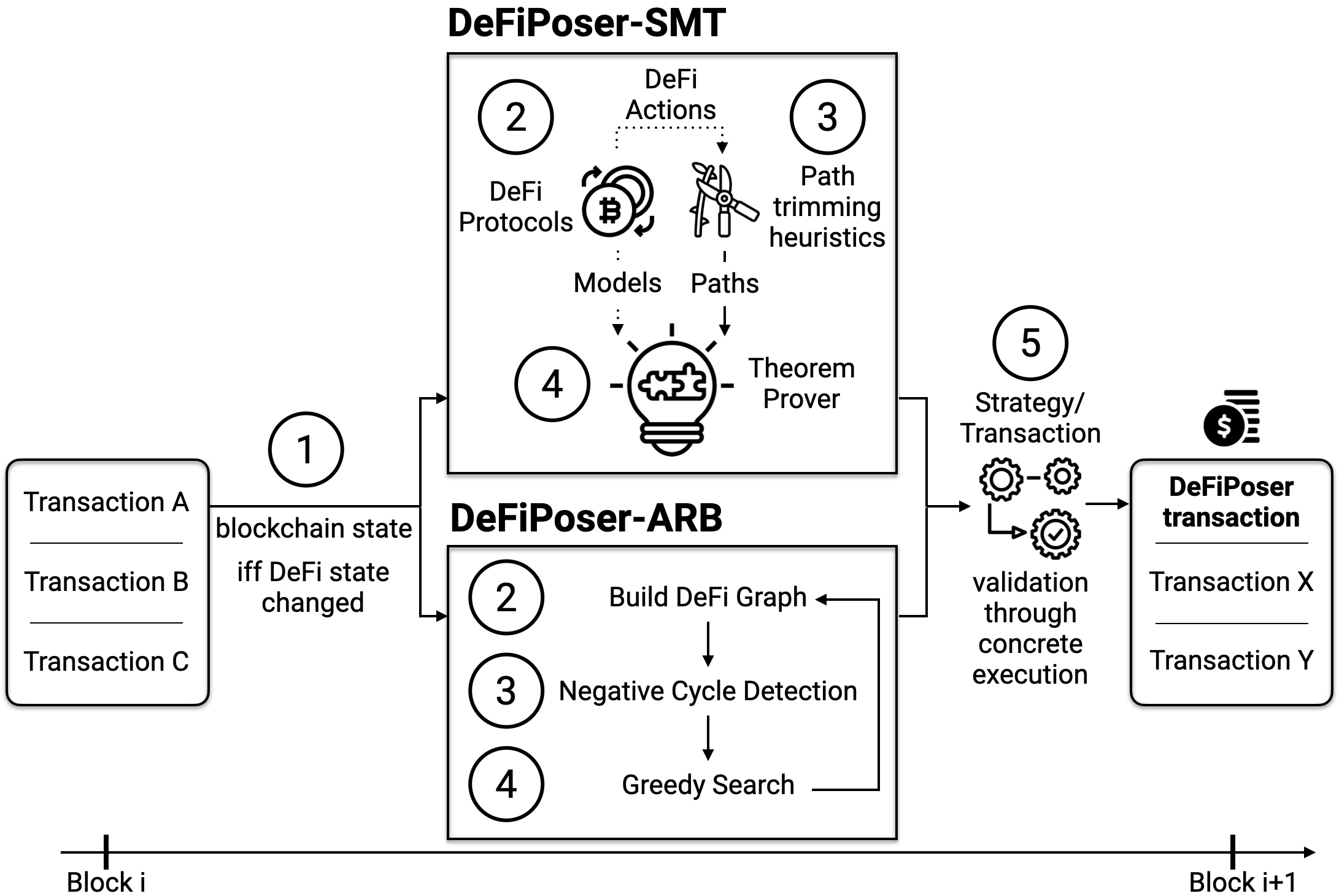}
\caption{\toolBF and \toolZThree system overview. In~\toolZThree, we \circled{2} Create logical models, \circled{3} paths are created and trimmed with heuristics and \circled{4} used within a theorem prover to generate a transaction. In \toolBF we \circled{2} build a graph of the blockchain state, \circled{3} identify negative cycles, \circled{4} perform a local search and repeat. The transaction with the highest revenue is \circled{4} concretely evaluated before being mined in the next block.}
\label{fig:model}
\end{figure}


A peculiarity of DeFi platforms is their ability to inter-operate; e.g., one may borrow a cryptocurrency asset on one platform, exchange the asset on another, and for instance, lend the resulting asset on a third system. DeFi's \emph{composability} has led to the emergence of chained trading and arbitrage opportunities throughout the tightly intertwined DeFi space. Reasoning about what this easy composition entails is not particularly simple; on one side, atomic composition allows to perform \emph{risk-free arbitrage}~---~that is to equate asset prices on different DeFi markets. Arbitrage is a benign and important endeavor to keep markets synchronized.

On the other side, we have seen multi-million-revenue trades that cleverly use the technique of flash loans to exploit \emph{economic states} in DeFi protocols (e.g., the economic attack on bZx~\cite{bzxnetwork, qin2020attacking} Harvest Finance~\cite{harvest-post-mortem}, Value Defi~\cite{value-defi-post-mortem} and others~\cite{ akropolis-post-mortem, origin-dollar-post-mortem}). 
While exploiting economic states, however, is not a security attack in the traditional sense, the practitioners' community often frames these high-revenue trades as ``hacks.'' Yet, the executing trader follows the rules set forth by the deployed smart contracts. Irrespective of the framing, liquidity providers engaging with DeFi experience millions of USD in unexpected losses. This highlights the need for automated tools that help protocol designers and liquidity providers to understand arbitrage and financial implications in general when engaging with DeFi protocols.



\point{\toolBF and \toolZThree}
This paper presents two tools (cf.\ Figure~\ref{fig:model}) that automatically create transactions to compose existing DeFi protocols to generate revenue that can be extracted from the Ethereum ecosystem. They are designed to run in real-time: at every block, they can find (and execute) a new profit-generating transaction; we show how our running time of an unoptimized implementation requires an average of~\AverageAnalsysTimePerBlockBF and~\AverageAnalsysTimePerBlockZThree on a recent Ethereum block (for \toolBF and \toolZThree respectively), which is below Ethereum's average block time of~\AverageBlockTime~\cite{ethereum-block-time}. We would like to point out that \toolBF and \toolZThree, are best-effort tools: because the state of the blockchain and DeFi platforms may change at each block, it is important to operate in real-time, otherwise found trading opportunities might be outdated. Therefore, we made the choice of prioritizing execution speed over completeness, and we do not claim to find optimal strategies.

To the best of our knowledge, we are the first to provide automated transaction search mechanisms for composable DeFi protocols. The main risks for a trader using the tools that we consider within this work are currency exposure (i.e., price volatility risks) and the blockchain transaction fees. We discover that significant revenue can be generated with less than~$1$ ETH of initial capital when using flash loans.


\point{Our contributions are as follows}
\begin{itemize}
    \item \textbf{\toolBF:} We build a directed DeFi market graph and identify negative cycles with the Bellman-Ford-Moore algorithm. A local search then allows us to discover parameters for profitable arbitrage transactions in near-real-time (average of~\AverageAnalsysTimePerBlockBF per block).
    \item \textbf{\toolZThree and Space Reduction:} To discover more demanding trades than arbitrage, we model the DeFi systems using a state transition model, which we translate to a logical representation in the Z3 theorem prover. 
    We introduce heuristics to significantly prune the search space to achieve a near real-time transaction discovery (average of~\AverageAnalsysTimePerBlockZThree per block).
    \item \textbf{Miner Extractable Value (MEV) and Security:}
    We show how \toolZThree discovers the economic attack on bZx, which yields over~$0.48$M USD, and that this opportunity window was open for over~$69$ days. Given optimal adversarial mining strategies provided by a Markov Decision Process, we show quantitatively that MEV opportunities can deteriorate the blockchain security. For example, a rational MEV-aware miner with a hash rate of~$10\%$ will fork the blockchain if an MEV opportunity exceeds~$4$ times the block reward and the miner failed to claim the source of MEV.
    \item \textbf{Trading Strategy Validation:}
    We validate the trading strategies discovered by \toolBF and \toolZThree on a locally-deployed blockchain that mirrors the real network. We estimate that the found strategies yield~\TotalRevenueConfirmedBF and \TotalRevenueConfirmedZThree of profit between the Ethereum block~\StartingBlock to~\EndingBlock (\NumDays from December~2019 to May~2020). We demonstrate that our tools' capital requirements are minimal: the majority of the strategies require less than~\CapitalRequrementWithoutFlashLoan, and only~\CapitalRequrementWithFlashLoan when using flash loans.
\end{itemize}

\point{Paper organization}
The remainder of the paper is organized as follows.  Section~\ref{sec:background} elaborates on the DeFi background, discusses stable coins and flash loans. 
Section~\ref{sec:defi_modeling} describes how we encode DeFi protocols into state transition models.
Section~\ref{sec:defiposerbf} applies negative cycle detection to find DeFi arbitrage opportunities.
Section~\ref{sec:technique} presents our heuristics and techniques to enable the autonomous discovery of adversarial strategies. 
Section~\ref{sec:experimental_evaluation} presents our empirical evaluation and quantitative analysis of the found strategies on previous Ethereum blockchain blocks.
Section~\ref{sec:security} discusses \tool's blockchain security implication.
We discuss related works in Section~\ref{sec:relatedwork} and conclude the paper in Section~\ref{sec:conclusion}.

\section{Background}
\label{sec:background}
In this section, we outline the required background for DeFi. For extensive background on blockchains and smart contracts, we refer the interested reader to~\cite{bonneau2015sok, atzei2017survey}.

\subsection{Decentralized Finance (DeFi)}\label{sec:defi-platform-background}
Decentralized Finance~(DeFi) refers to a financial ecosystem that is built on top of (permissionless) blockchains~\cite{wust2018you}. DeFi supports a multitude of different financial applications~\cite{dydx, uniswap2018,makerdao, compoundfinance, dydx, bzxnetwork, compoundfinance, aave, dydx, qin2020attacking}. The current DeFi landscape is mostly built upon smart contract enabled blockchains (e.g., Ethereum). We briefly summarize relevant DeFi platforms.

\point{Automated Market Maker~(AMM)} In traditional finance, asset exchanges are usually operated in the form of order matching. Asks and bids are matched in a centralized limit order book, typically following the FIFO principle \cite{chen_2020}. In DeFi, such an order matching mechanism would be inefficient because the number of transactions per second supported by the underlying blockchain is usually limited. Therefore, AMM minimizes the number of transactions required to balance an on-chain asset exchange. AMM allows liquidity providers, the traders who are willing to provide liquidity to the market, to deposit assets into a liquidity pool. Liquidity takers then directly trade against the AMM liquidity pool according to a predefined pricing mechanism. The constant product AMM is currently the most common model (adopted by over~$66\%$ of the AMM DEX), where the core idea is to keep the product of the asset amounts in the liquidity pool constant. Consider a constant product AMM that trades the asset pair $X$/$Y$. $x$ and $y$ are the amount of $X$ and $Y$ respectively in the liquidity pool. A liquidity taker attempts to sell $\Delta x$ of $X$ and get $\Delta y$ of $Y$ in exchange. The constant product rule stipulates that $x\times y = (x+\Delta x)\times(y-\Delta y)$. Uniswap~\cite{uniswap2018} is the most dominating constant product AMM with a market capitalization of~$1.4$B USD~\cite{defipulse}. Variant AMMs utilize different pricing formulas, e.g., Bancor~\cite{hertzog2017bancor}, while other platforms (e.g., Kyber~\cite{kyber}) aggregate AMMs. When receiving an order from a user, these platforms redirect the order to the AMM, which provides the best asset price.

\point{Stablecoin} 
Stablecoins are a class of cryptocurrencies designed to alleviate the blockchain price volatility~\cite{mita2019stablecoin}. The most salient solution for stabilization is to peg the price of stablecoins to a less-volatile currency (e.g., USD)~\cite{moin2020sok}. There exist over~$200$ stablecoin projects announced since~2014~\cite{HowManyS88:online}. Among them, SAI and DAI developed by MakerDAO~\cite{makerdao} have received extensive attention. Both SAI and DAI are collateral-backed stablecoins. SAI is collateralized solely by ETH, whereas DAI is an SAI upgrade to support multiple assets as collateral. At the time of writing, the collateral locked in MakerDAO amounts to~$2.73$B USD~\cite{defipulse}.

\point{Flash Loans} 
The Ethereum blockchain operates similarly to a replicated state machine. Transactions trigger state transitions and provide the input data necessary for the Ethereum Virtual Machine (EVM) state to change according to rules set by smart contracts. Interestingly, the EVM state is only affected by a transaction if the transaction executes without failure. In the case of a failed transaction, the EVM state is reverted to the previous state, but the transaction fees are still paid to miners (as in to avoid Denial of Service attacks). A transaction can fail due to the following three reasons: Either the transaction sender did not specify a sufficient amount of transaction fees, or the transaction does not meet a condition set forth by the interacting smart contract, or the transaction is conflicting (e.g., double-spending) with another transaction.

This concept of a state reversion enables the introduction of flash loans, short-lived loans that execute atomically within only one blockchain transaction. Within a single transaction, (i) the loan is taken from a liquidity pool, (ii) the loan is put to use, and (iii) the loan (plus interest payment) is paid back to the flash loan pool. If the third condition is not met, i.e., the loan plus interests are not paid back, then the entire flash loan transaction fails. This is equivalent to the case that the loan was never issued because the EVM state is not modified out of the result of a failed transaction. 

Flash loans, therefore, entail two interesting properties. First, the lender is guaranteed that the borrower will repay the loan. If the repayment is not performed, the loan would not be given. Second, the borrower can technically request any amount of capital, up to the amount of funds available in a flash loan pool, given a constant payment which corresponds to the blockchain transaction fees (about~\empirical{10 USD}\xspace for the most common flash loan providers). The borrower hence can have access to millions of USD with just a few initial USD and hence is not exposed to the currency risk of the lent asset.

\section{DeFi Modeling}
\label{sec:defi_modeling}
We proceed to introduce our system, trader, and state transition model for the interaction between DeFi platforms. On a high level, our model state consists of the DeFi market states, as well as the cryptocurrency asset balances of a trader $\mathbb{T}$. The transitions represent DeFi actions performed by the trader $\mathbb{T}$ on the respective DeFi platforms. The goal of the trader is to maximize the amount of cryptocurrency assets held.

\subsection{System Model}
\begin{figure}[tb]
\centering
\includegraphics[width = \columnwidth]{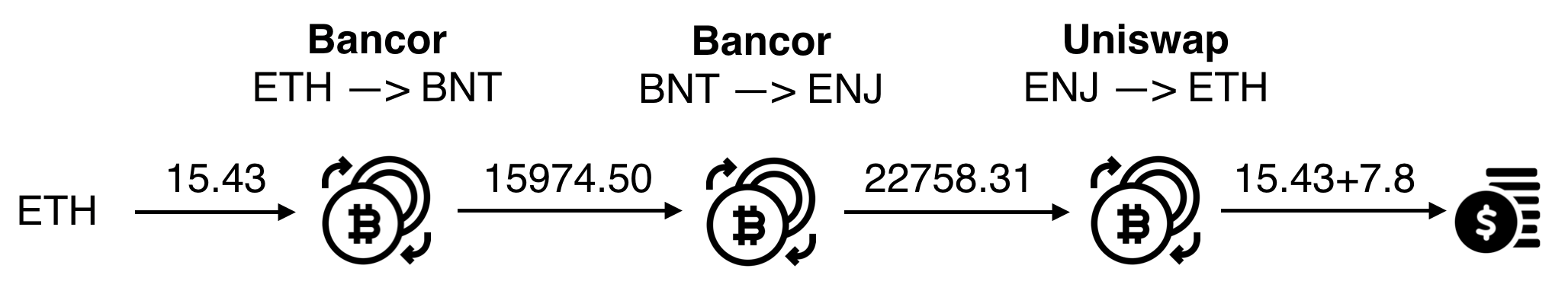}
\caption{Example strategy across three DeFi markets, identified at Ethereum block $10,001,087$, which would yield a revenue of \Ether{7.81}.}
\label{fig:example-most-revenue}
\end{figure}

Our system consists of a blockchain with financial cryptocurrency assets (i.e., coins or tokens). Cryptocurrency assets can be used within DeFi platforms (i.e., markets), such as exchanges, lending, and borrowing platforms. Each DeFi platform offers a set of \emph{actions}, which can be triggered by a transaction. Actions take an asset as input and yield, for instance, another asset as output. Multiple actions can be encapsulated in one transaction and executed atomically in sequence. A \emph{path}, is a sequence of actions across DeFi platforms. We denote as \emph{strategy} (cf.\ Figure~\ref{fig:example-most-revenue}), or transaction, a path with parameters for each action (such as coin amounts, etc.). We consider a state of a DeFi market to change whenever an action manipulates the amount of assets within this DeFi market. Note that we only consider the blockchain state at block-height $i$ after the execution of all transactions within a block $i$ (i.e., we do not consider intermittent block states).

\subsection{Trader Model}
\label{sec:threat_model}
We consider a computationally bounded trader (denoted by $\mathbb{T}$) which is capable of executing transactions (i.e., perform actions) across a set of DeFi platforms. $\mathbb{T}$'s cryptocurrency assets are limited by the supply of liquidity available in public flash loan pools~\cite{qin2020attacking}. The trader is capable of reading the blockchain contents but is not expected to observe unconfirmed blockchain transactions on the network layer. We assume that the trader is capable of placing a transaction ahead of other DeFi transactions within a future blockchain block. Practically, this requires the trader to pay a higher transaction fee, as most miners appear to order transactions based on gas price. We assume that the trader is not colluding with a miner, while this may present an interesting avenue for future work.

We assume that the trader is operating on the blockchain head, i.e., the most recently mined, valid block, of the respective blockchain. In the case of a Proof-of-Work (PoW) blockchain, the most recent block shall also be the one with the most PoW (i.e., the greatest difficulty). For simplicity, we ignore complications resulting from blockchain forks.

\subsection{Notation} 
\label{sec:notation}
To ease the understanding of the following paragraphs, we proceed by introducing the utilized notation.
\point{Assets} The set $C$ denotes the collection of cryptocurrency assets, which the trader uses to generate trading strategies.
\point{Actions} The set $A$ denotes the collection of actions the trader selects from the DeFi protocols.
\point{Parameters} The trader $\mathbb{T}$ must supply parameters to execute actions $a \in A$, e.g., the amounts of cryptocurrency assets $\mathbb{T}$ sends to the corresponding DeFi platforms.
\point{Path} A path $p \in P$ is a sequence of $n$ non-repeated actions drawn from $A$.
We denote the power set of all actions with $\powerset(A)$, which consists of all subsets of the action set $A$, including the empty set.
Given a subset $K \in \powerset(A)$, we denote the permutations set of $K$ with $\mathfrak{S}(K)$.
The collection of all paths $P$ can then be defined using Equation~\ref{eq:path}.
Note $P$ consists of paths of different lengths.
\begin{equation}\label{eq:path}
\begin{aligned}
P = \cup^{\powerset(A)}_{K} \mathfrak{S}(K),\quad \text{s.t.} \quad & \forall p = (a_1, a_2, ..., a_n) \in P\\
                                                               & a_i \in A, \forall i \in [1, n] \\
                                                               & a_i \neq a_j, \forall a_i, a_j \in A, i \neq j
\end{aligned}
\end{equation}
\point{Strategy} A strategy consists a path $p \in P$ with $n$ actions, a list of parameters $[x_1, \ldots, x_n]$ for each action in $p$, and an initial state (cf. Equation~\ref{eq:state}) of the model.
\point{Balance function} Given a strategy with $n$ actions, the balance function $\mathcal{B}^\mathbb{T}_i(c)$ denotes $\mathbb{T}$'s balance for cryptocurrency asset $c$ after performing the $i^\mathit{th}$ action, where $0 \leq i \leq n$ and $c \in C$.
\point{Storage function} $\mathcal{K}(a)$ denotes the set of smart contract storage variable addresses an action $a$ reads from and writes to. These addresses are identified from the underlying blockchain runtime environment. We use $\mathcal{K}^{\mathbb{T}}(a)$ to denote a subset of $\mathcal{K}(a)$, which is only relevant to the trader $\mathbb{T}$. 

\subsection{States}
We classify the state variables into two categories, the trader and DeFi states. $S^{\mathbb{T}}$ represents the trader's asset portfolio (cf.\ Equation~\ref{eq:adversarial_state}). $S^{\text{DeFi}}$ is the set of all storage variables $\mathbb{T}$ reads from and writes to, for all the DeFi actions in our model (cf.\ Equation~\ref{eq:defi_state}). The union $S$ of these two categories is the overall state of our system (cf.\ Equation~\ref{eq:state}). Given a strategy with $n$ actions, the state after performing the $i^\mathit{th}$ action, where~$0 \leq i \leq n$ is denoted as $s_i$, with the initial state $s_0$.
\begin{equation}\label{eq:adversarial_state}
S^{\mathbb{T}} = \{\mathcal{B}^\mathbb{T}(c) : \forall c \in C\}
\end{equation}
\begin{equation}\label{eq:defi_state}
S^{\text{DeFi}} = \cup_{\forall a \in A} \mathcal{K}^\mathbb{T}(a)
\end{equation}
\begin{equation}\label{eq:state}
    S = S^{\mathbb{T}} \cup S^{\text{DeFi}}
\end{equation}

\subsection{Transitions}
Our state transition function is $\mathcal{F}^{\mathbb{T}}(s \in S, a \in A, x) \rightarrow S$, outputs the next state if action $a$ with parameter $x$ is performed on state $s$ by trader $\mathbb{T}$. Given a strategy with $n$ actions, where $a_i$ and $x_i$ represents the $i^\mathit{th}$ action and parameter respectively, and $s_i$ represents the state after the $i^\mathit{th}$  action. 
Equation~\ref{eq:transition} shows the state transition process of this strategy, while Equation~\ref{eq:final_state} computes the final state $s_n$ when each action is sequentially applied to $s_0$.
\begin{equation}\label{eq:transition}
    s_{i + 1} = \mathcal{F}^{\mathbb{T}}(s_{i}, a_{i+1}, x_{i+1})
\end{equation}
\begin{equation}\label{eq:final_state}
s_n = \mathcal{F}^{\mathbb{T}}(\ldots \mathcal{F}^{\mathbb{T}}(\mathcal{F}^{\mathbb{T}}(s_0, a_1, x_1), a_2, x_2) \ldots)
\end{equation}

\subsection{Objective}
We choose an asset $b \in C$ as our \emph{base} cryptocurrency asset. The objective of the trader $\mathbb{T}$ is to find a strategy, such that the balance of $b$ (cf. Equation~\ref{eq:obj}) is maximized, whereas the portfolio balances of the trader, except for $b$, remain the same.
\begin{equation}\label{eq:obj}
\begin{aligned}
\text{maximise}_{p \in P}\ \text{obj}(s_0, p) = \mathcal{B}^\mathbb{T}_n(b) - \mathcal{B}^\mathbb{T}_0(b) \\
\text{with constraints:}\ \mathcal{B}^\mathbb{T}_n(c) = \mathcal{B}^\mathbb{T}_i(c), \forall c\in C \setminus b
\end{aligned}
\end{equation}

\subsection{\emph{Base} cryptocurrency asset}
To identify revenue yielding paths, we make the assumption that the trader $\mathbb{T}$ operates in this work on a single \emph{base} cryptocurrency asset. Naturally, this can be extended to multiple \emph{base} currencies to increase potential financial results.

\subsection{\tool Design Choices}
\label{sec:desgin_choices}

\begin{figure}[tb]
\begin{center}
\includegraphics[width = \columnwidth]{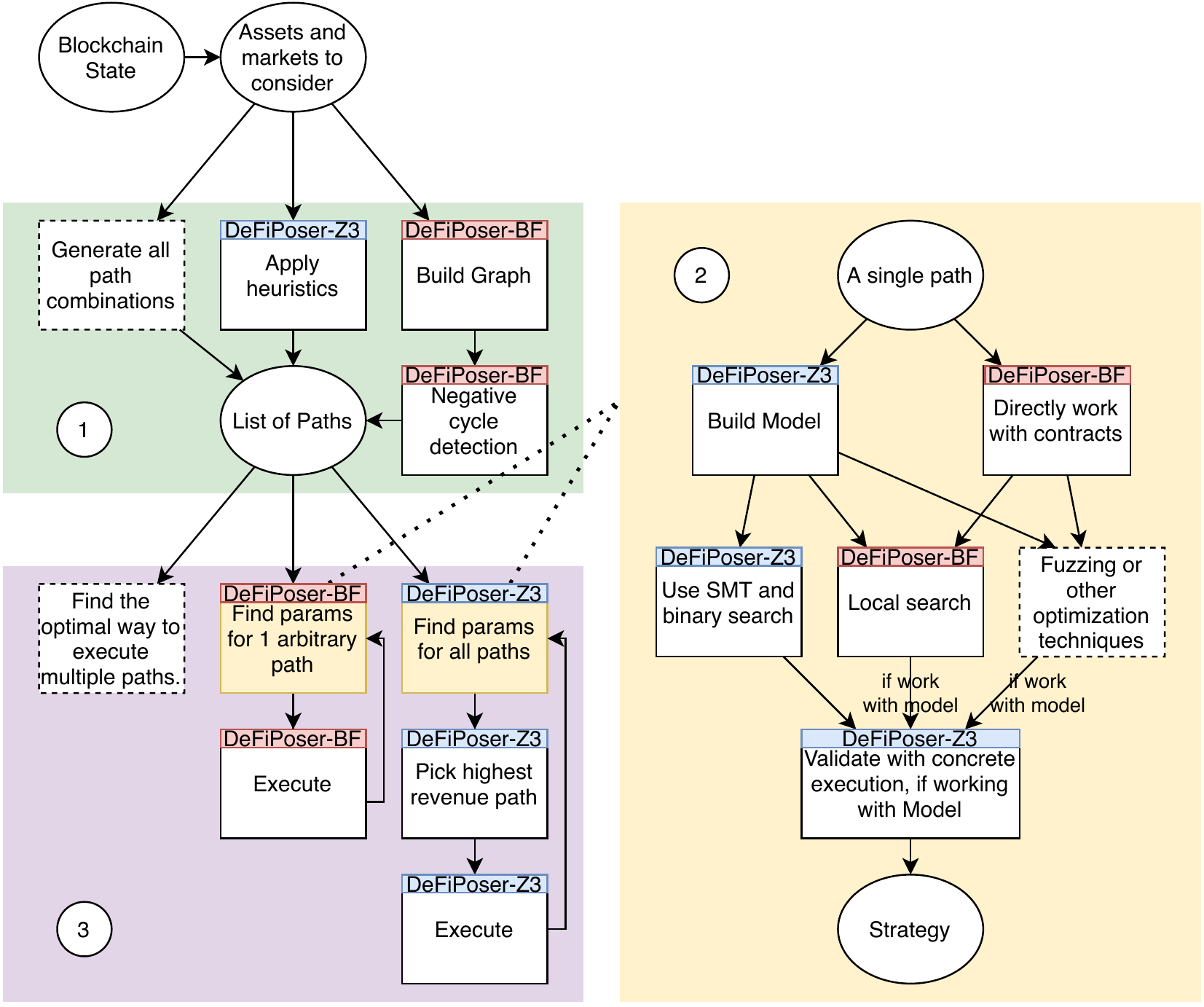}
\end{center}
\caption{Technical design choices of \tool. \tool consists of three components: \circled{1} a path pruning component; \circled{2} a parameter search component, and \circled{3} a strategy combination/execution component.}
\label{fig:design_chocie}
\end{figure}

Figure~\ref{fig:design_chocie} shows the high-level design choices of the \tool tools we present in this paper. \tool consists of three components: \circled{1}, a pruning algorithm to filter potentially profitable paths; \circled{2}, a search algorithm which searches parameters to maximize the revenue of a given path, and \circled{3} a strategy combination/execution algorithm, which decides how the found strategies are executed.

Generally speaking, each instantiation of the different components bears its own advantages and disadvantages. For instance, negative cycle detection only searches cyclic paths, whereas pruning with heuristics can search for any path structure. Given a simple path such as a cyclic arbitrage, we find that local search is faster than the SMT solver (cf.\ Figure~\ref{fig:outliers}), but does not provide satisfiability proofs. In the following we present two variants of \tool, namely \toolBF (cf.\ Section~\ref{sec:defiposerbf}) and \toolZThree (cf.\ Section~\ref{sec:technique}).

\section{Applying Negative Cycle Detection to DeFi Arbitrage}
\label{sec:defiposerbf}
Previous works propose negative cycle detection algorithms, such as the Bellman-Ford-Moore algorithm, to find arbitrage opportunities\cite{cherkassky1999negative}. In these algorithms, the exchange markets are modeled as a directed weighted graph ($g$). Every negative cycle in the graph then corresponds to an arbitrage opportunity.

\subsection{Negative Cycle Detection to Detect Arbitrage}
We adopt the following notations to translate arbitrage detection into a negative cycle detection problem.

\point{Nodes} The set $N$ denotes the collection of nodes. Each node (vertex) represents a different asset ($c \in C$).

\point{Directed edges} The set $E$ denotes the collection of all edges. An edge $e_{i,j}$ that points from asset $c_i$ to $c_j$ represents that there exist a market where the trader $\mathbb{T}$ can sell cryptocurrency asset $c_i$ to purchase cryptocurrency asset $c_j$.

\point{Spot price} The spot price $p^{\text{spot}}_{i,j}$ for edge $e_{i,j}$ is the approximated best current price a trader $\mathbb{T}$ finds on all DeFi AMM markets, when selling an arbitrarily small amount (close to $0$) of a cryptocurrency asset $c_i$ to purchase $c_j$. 
\point{Arbitrage} A path  $[c_1 \xrightarrow{a_1} c_2 \ldots c_{k-1} \xrightarrow{a_{k-1}} c_{k}]$ consists an arbitrage opportunity, if $p^{\text{spot}}_{1,2} \times \ldots \times p^{\text{spot}}_{k-1,k} > 1$.

\point{Edge weight} To apply negative cycle detection algorithms, we use the negative log of price $w_{i,j} = -log(p^{\text{spot}}_{i,j})$ as the weights for edge $e_{i, j}$. An arbitrage opportunity exists if $w_{1,2} + \ldots + w_{i-1,i} < 0$. 

\point{Path finding} Our objective is to maximize $\mathbb{T}$'s \emph{base} cryptocurrency asset. An arbitrage cycle, however, may not consist of the \emph{base} asset. Therefore, we convert the arbitrage revenue to the \emph{base} cryptocurrency asset by the end of the execution. More concretely, we find all `connecting' markets that support the conversion between one of the arbitrage assets and the \emph{base} cryptocurrency asset. We perform the conversion using the `connecting' markets with the best price.

\subsection{Negative Cycle Detection Algorithms}


\begin{algorithm}[]
\SetAlgoLined
\DontPrintSemicolon
\SetKwProg{Fn}{Function}{ is}{end}
\KwIn{\;
$s_0$ $\gets$ Initial state $;\,$ \textit{target} $\gets$ Minimum revenue target\;
}
\KwOut{$\textit{revenue}_{total}$}
\textit{s} $\gets s_0$ $;\,$ \textit{g} $\gets$ buildGraph($N$, $E$, $s$) $;\,$ $\textit{revenue}_{total}$ $\gets$ 0\;
\While{hasNegativeCycle(g)}{
    \textit{cycle} $\gets$ getNegativeCycle(\textit{graph})\;
    \textit{p} $\gets$ getPath(\textit{cycle})\;
    (\textit{revenue}, \textit{s}) $\gets$ search(\textit{p})\;
    \If{\textit{revenue} > target}{
        $\textit{revenue}_{total}$ $\gets$ $\textit{revenue}_{total}$ $+$ \textit{revenue}\;
    }
    \textit{g} $\gets$ buildGraph($N$, $E$, $s$)\;
}
\Return{$\textit{revenue}_{total}$}\;
\;
\Fn{buildGraph(N, E, $s \in S$)}{
    $\text{\# fetch the spot price for each } e \in E$\;
    $\text{\# build the graph } \textit{g}; \text{where } w_{c_i, c_j} = -log(p_{c_i, c_j}^{\text{spot}})$\;
    return \textit{g} \;
}
\Fn{hasNegativeCycle(\textit{g})}{
    return \textit{(Detects a negative cycle?)} \;
}
\Fn{getPath(\textit{cycle})}{
    $p \in P \text{ connects } \mathbb{T} \text{'s baseasset with } \textit{cycle.}$\;
    return \textit{p} \;
}
\Fn{search(\textit{p})}{
    $\text{\# find the parameters for path } \textit{p}$\;
    \textit{s'} $\gets$ \text{state after executinng the strategy}\;
    return \textit{(\textit{revenue}, \textit{s'})}
}
\caption{Negative cycle arbitrage detection.}
\label{alg:bf}
\end{algorithm}

Negative cycle detection algorithms combine the shortest path algorithm with a cycle detection strategy. Cherkassky~\etal\ \cite{cherkassky1999negative} studied various combinations of shortest path algorithms (Bellman-Ford-Moore\cite{bellman1958routing,ford2015flows,moore1959shortest}, Goldfarb-Hao-Kai\cite{goldfarb1991shortest}, Goldberg-Radzik\cite{goldberg1993heuristic}, etc.) and cycle detection strategies (Walk to the root, Admissible graph search\cite{goldberg1995scaling}, Subtree traversal\cite{kennington1980algorithms}, etc.) and compared their relative performances. A natural question is whether these cycle detection algorithms can be directly applied to find profitable transactions in DeFi. 

In bid-ask markets, the price does not change if the trade volume is within the bid/ask size \cite{chen_2020}. DeFi AMM exchanges, however, follow a dynamic price based on the trade volume. Intuitively, the bigger the transaction size, the worse the trading price becomes. Hence, our algorithm needs to consider dynamic price changes and update the graph $g$ after every action. On a high level, a Bellman-Ford-Moore inspired algorithm repeatedly performs the following steps: (i) Build the graph $g$ based on the spot prices from the current state $s \in S$; (ii) Detect arbitrage cycles in the graph $g$ (Bellman-Ford-Moore); (iii) Build a path based on the negative cycle, and find the strategy (parameters for the path), finally (iv) Execute the strategy and update the state $s$. Algorithm~\ref{alg:bf} presents the details of \toolBF. To find the parameters for a path, Algorithm ~\ref{alg:bf} gradually increases the amount of \emph{base} assets into the path until there is no increase in revenue. 

We present \toolBF's evaluation in Section~\ref{sec:experimental_evaluation}.


\section{Design of \toolZThree}
\label{sec:technique}
\begin{figure*}[htb!]
     \centering
     \begin{subfigure}[b]{0.3\textwidth}
         \centering
         \includegraphics[width=\textwidth]{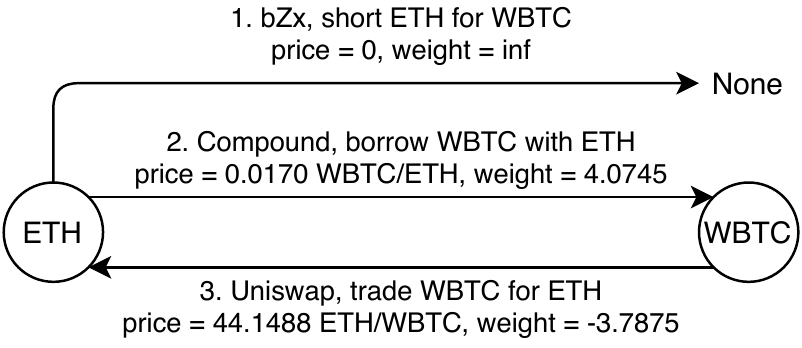}
         \caption{Initial state graph. \\ Cycle weight sum $= 4.07 - 3.79 = 0.28$}
         \label{fig:bf_1}
     \end{subfigure}
     \hfill
     \begin{subfigure}[b]{0.3\textwidth}
         \centering
         \includegraphics[width=\textwidth]{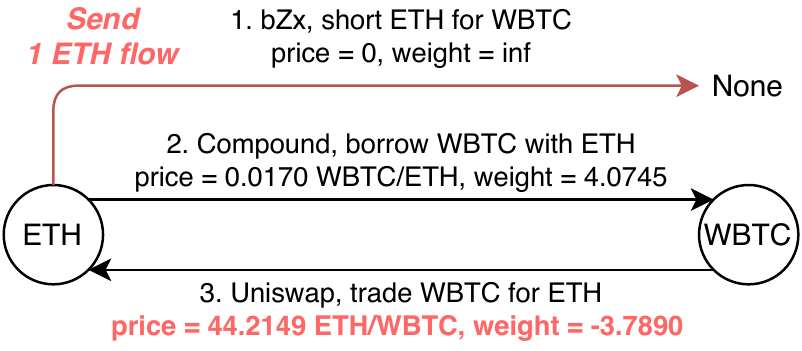}
         \caption{Graph after shorting $1$ ETH. \\ Cycle weight sum $= 4.07 - 3.79 = 0.28$}
         \label{fig:bf_2}
     \end{subfigure}
     \hfill
     \begin{subfigure}[b]{0.3\textwidth}
         \centering
         \includegraphics[width=\textwidth]{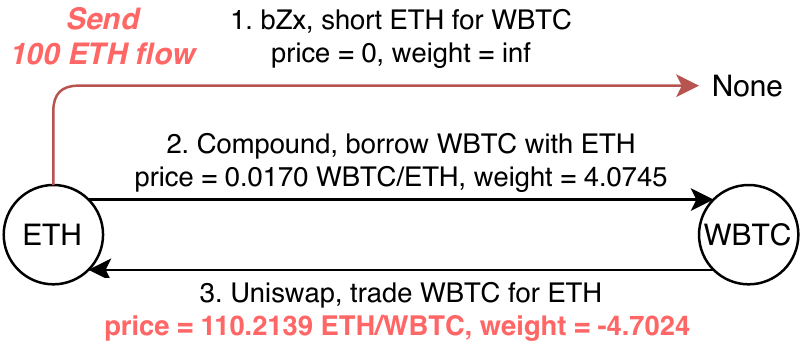}
         \caption{Graph after shorting $1,000$ ETH. \\ Cycle weight sum $= 4.07 - 4.70 = -0.63$}
         \label{fig:bf_3}
     \end{subfigure}
     \caption{Directed weighted graph for the economic bZx attack on the Ethereum block $9,462,687$. Shorting ETH for WBTC on bZx does not return assets to the trader $\mathbb{T}$, and the action, therefore, does not point to any cryptocurrency assets. Graph~\ref{fig:bf_1} has no arbitrage opportunity on the WBTC/ETH market ($0.0170 \times 44.1488 = 0.75 \leq 1$). In Graph~\ref{fig:bf_2} and~\ref{fig:bf_3}, the weights change (ETH/WBTC price) after the trader increases the flow (in ETH) to the bZx market because bZx's price depends on the Uniswap price. The graph is hence dynamic~\cite{chandrachoodan2001adaptive}, i.e., the weights need to be updated after each action. The action encoding of \toolZThree models the bZx's price dependence on Uniswap. Note that the bZx attack does not violate \textbf{Heuristic 6}, because action $1$ does not return any asset nor forms a sub-path (cf.\ Figure~\ref{fig:asset_dependency}).}
     \label{fig:bf}
\end{figure*}

In this section, we discuss an alternative technique, \toolZThree, to find profitable transactions in DeFi, which is more general when compared to \toolBF. More specifically, \toolZThree can operate on non-cyclic strategies, while \toolBF cannot. We observe that profitable DeFi strategies do not necessarily form a complete cycle. For example, Figure~\ref{fig:bf_1} shows the graph for the economic bZx attack (cf.\ Section~\ref{sec:security}). The strategy requires the trader to send Ether to edge $1$ without receiving any assets in return and to then perform an arbitrage cycle with edge $2$ and $3$.

\subsection{Choosing an SMT Solver for \toolZThree}
To overcome the aforementioned challenges of non-existent cycles, we chose to adopt a theorem prover for \toolZThree's (cf.~Figure~\ref{fig:model}) design. The theorem prover logically formulates what a profitable strategy entails to locate concrete profitable instantiations. We perform systematic path exploration to determine if the model (cf. Section~\ref{sec:defi_modeling}) satisfies the provided requirements, similar to other model checking systems~\cite{tsankov2018securify, brent2018vandal, grech2018madmax, luu2016making, krupp2018teether, chang2019scompile, kalra2018zeus, tikhomirov2018smartcheck}.

Our model requires the SMT solver (such as MathSat~\cite{bruttomesso2008mathsat}, Z3~\cite{de2008z3}, or Coral~\cite{souza2011coral}) to support floating-point arithmetic because we adopt the theory of real numbers (cf. Section~\ref{sec:defi_modeling}). We encode the state transition model in three major steps: (i) Encode the initial state as a predicate; (ii) iteratively apply state transition actions, and encode the resulting states after each action as predicates. Then, (iii) convert the objective function into a set of constraints to ensure that the value of the trader portfolio increases by $Z$, and translate the constraints into predicates. Note that we rely on an optimization algorithm (cf. Algorithm~\ref{alg:optimiser}) to find the highest possible $Z$. The optimization process requires solving the same SMT problem with different initializations of $Z$ (cf.\ Appendix~\ref{app:encoding_example} for an example).

\subsection{Path Pruning}
\label{sec:pruning}
One bottleneck of model checking is the combinatorial path explosion problem. We, therefore, prune the paths by applying the following heuristics. Note that heuristics may prune profitable strategies, and \toolZThree is therefore only a best-effort tool.


\point{Heuristic 1} A profitable strategy must consist of more than one action. That is because, given an initial state $S_0$, a strategy with only one action will not increase the balance of the \emph{base} cryptocurrency asset while keeping the balance of all other cryptocurrency assets unchanged.

\point{Heuristic 2} A strategy must start with a sequence of entering actions. An entering action is defined as any action which takes the \emph{base} cryptocurrency asset as input.

\point{Heuristic 3} A strategy must end with a sequence of exiting actions. An exiting action is defined as any action that outputs the \emph{base} cryptocurrency asset. Recall that the objective of the trader is to maximize the amount of \emph{base} assets held.

\point{Heuristic 4} Apart from the entering actions, an action must depend on at least one previous action. Conceptually, this is to avoid a strategy to contain actions that do not interact with any other actions. Given two actions $a_i, a_j \in A$, we define that $a_i$ and $a_j$ are independent actions, iff.\ there is no intersection between $\mathcal{K}^{\mathbb{T}}(a_i)$ and $\mathcal{K}^{\mathbb{T}}(a_j)$ (cf. Equation~\ref{eq:indepenent}). In other words, the execution of $a_i$ does not affect the execution results of $a_j$, no matter what concrete state is given.
\begin{equation}\label{eq:indepenent}
{a_i} \perp \!\!\! \perp {a_j} \iff \mathcal{K}^{\mathbb{T}}(a_i) \cap \mathcal{K}^{\mathbb{T}}(a_j) = \emptyset
\end{equation}
Recall that $\mathcal{K}(a)$ denotes the set of smart contract storage variables an action $a$ reads from and writes to, and $\mathcal{K}^{\mathbb{T}}(a)$ denotes a subset of $\mathcal{K}(a)$, which is relevant to the trader $\mathbb{T}$. As an example of independence, we assume $a_1$ transacts $c_1$ to $c_2$ using a constant product market $M1$ with liquidity $L1^{c_1}$ and $L1^{c_2}$, and $a_2$ transacts $c_1$ to $c_3$ using another constant product market $M2$ with liquidity $L2^{c_1}$ and $L2^{c_3}$. Equation~\ref{eq:storage_dependency} shows the storage variables $a_1$ and $a_2$ reads from and writes to. $a_1$ and $a_2$ are not independent, as they both read/write variable $\mathbb{T}.c_1$. Therefore, Heuristic 4 does not prune the path containing $a_1$ and $a_2$.
\begin{equation}
\begin{aligned}
\mathcal{K}^{\mathbb{T}}(a_1) = \{ M1.L1^{c_1},M1.L1^{c_2},\mathbb{T}.c_1,\mathbb{T}.c_2 \} \\
\mathcal{K}^{\mathbb{T}}(a_2) = \{ M2.L2^{c_1},M2.L2^{c_3},\mathbb{T}.c_1,\mathbb{T}.c_3 \}
\end{aligned}
\label{eq:storage_dependency}
\end{equation}
\point{Heuristic 5} An action cannot be immediately followed by another reversing action (i.e., a mirroring action) on the same DeFi market. For instance, if $a_1$ transacts $c_1$ to $c_2$, and $a_2$ converts $c_2$ to $c_1$ on the same market, then heuristic 5 will prune all paths that contain $a_1, a_2$.


\begin{figure}[tb]
\centering

\begin{subfigure}{\columnwidth}
\centering
\includegraphics[width = 0.7\columnwidth]{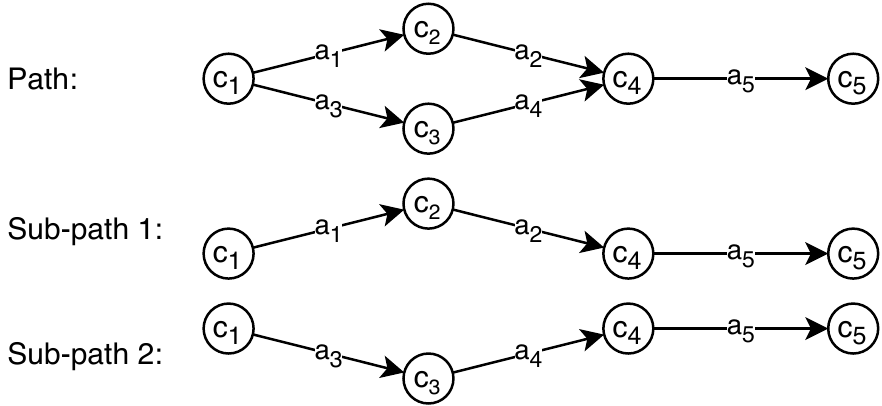}
\caption{Heuristic 6 - Branching example}
\label{fig:heuristic_6}
\end{subfigure}

\begin{subfigure}{\columnwidth}
\centering
\includegraphics[width = 0.7\columnwidth]{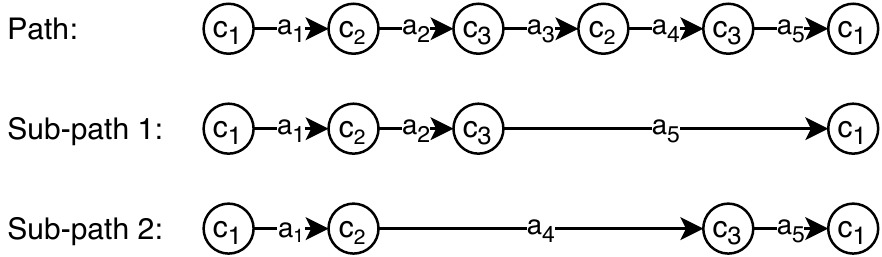}
\caption{Heuristic 7 - Looping example}
\label{fig:heuristic_7}
\end{subfigure}

\caption{Example of branching and looping paths.}
\end{figure}

\point{Heuristic 6} A path cannot include any branching. For example, a path of 5 actions $[c_1 \xrightarrow{a_1} c_2 \xrightarrow{a_2} c_4, c_1 \xrightarrow{a_3} c_3 \xrightarrow{a_4} c_4, c_4 \xrightarrow{a_5} c_1]$ is composed of two paths, $[c_1 \xrightarrow{a_1} c_2 \xrightarrow{a_2} c_4 \xrightarrow{a_5} c_1]$ and $[c_1 \xrightarrow{a_3} c_3 \xrightarrow{a_4} c_4 \xrightarrow{a_5} c_1]$ (cf.\ Figure~\ref{fig:heuristic_6}). In our work, we choose the more profitable path, and discard the other, because both paths affect the asset $c_4$. In a future work, it might be interesting to attempt to extract profit from both paths in an effort to maximize the revenue.

\point{Heuristic 7} A path must not include any loops. For example, a path $[c_1 \xrightarrow{a_1} c_2 \xrightarrow{a_2} c_3 \xrightarrow{a_3} c_2 \xrightarrow{a_4} c_3 \xrightarrow{a_5} c_1]$ consists of a loop between $c_2$ and $c_3$. This path is composed of two sub-paths, namely $[c_1 \xrightarrow{a_1} c_2 \xrightarrow{a_2} c_3 \xrightarrow{a_5} c_1]$ and $[c_1 \xrightarrow{a_1} c_2 \xrightarrow{a_4} c_3 \xrightarrow{a_5} c_1]$  (cf.\ Figure~\ref{fig:heuristic_7}). We again chose the more profitable path, and discard the other for simplicity. We leave it to future work to optimize the potential gain.

The efficiency of path pruning can be evaluated across two dimensions: \emph{(i)} the number of paths that are pruned, and, \emph{(ii)} the reduction in revenue resulting from the heuristic pruning. To address the former, we show the reduction of the number of paths due to the heuristics in Table~\ref{tab:heuristics} and discuss these results further in Section~\ref{sec:complexity}. Regarding the latter, because we cannot quantify the optimal revenue due to the combinatorial explosion of the search space, we, unfortunately, see no avenue to quantify the reduction in revenue caused by the heuristics.

\subsection{\toolZThree Revenue Optimizer}
\label{sec:optimization}
SMT solvers validate if any initialization of the free variables would satisfy the requirements defined. One requirement we specify is to increase the \emph{base} cryptocurrency asset balance by a fixed amount. To find the maximum satisfiable revenue, we chose to use the following optimization algorithm (cf.~Algorithm~\ref{alg:optimiser}). At a high level, to identify a coarse upper and lower revenue bound, this algorithm first attempts to solve, given multiples of~10 for the trader revenue. Given these bounds, we perform a binary search to find the optimal value.

\subsection{Comparing \toolZThree to \toolBF}
Table~\ref{tab:comparison} summarizes our comparison between \toolZThree and \toolBF. While arbitrage opportunities appear plentiful, \toolBF cannot capture non-cyclic transactions such as the bZx case. Because \toolZThree can encode any arbitrary strategy as an SMT problem, we argue that it is a more generic tool, as long as the underlying SMT solver can find a solution fast enough. We would like to stress again that both tools \toolBF and \toolZThree do not provide optimal solutions. \toolBF greedily searches for arbitrage and extracts revenue as each opportunity arises. To show that \toolBF does not find optimal solutions, we provide the following example at block $9,819,643$. Here, \toolZThree finds two opportunities: \begin{description}
\item[Strategy 1] $[ETH \xrightarrow{Bancor} BNT \xrightarrow{Bancor} MKR \xrightarrow{Uniswap} ETH]$ with $0.20$ ETH of revenue.
\item[Strategy 2] $[ETH \xrightarrow{Uniswap} BAT \xrightarrow{Bancor} BNT \xrightarrow{Bancor} MKR \xrightarrow{Uniswap} ETH]$ with $0.11$ ETH of revenue.
\end{description}
\toolZThree will only execute strategy 1. \toolBF, however, finds and executes strategy 2 first to extract $0.11$ ETH. After executing strategy 2 and updating the graph, strategy 1 is no longer profitable. Therefore, \toolBF only extracts a revenue of $0.11$ ETH in this block. Note that \toolZThree provides proof of satisfiable/unsatisfiable revenue targets for each considered path. However, \toolZThree remains a best-effort tool because the heuristics prune paths that may be profitable. Contrary to \toolBF, \toolZThree does not merge paths.

\begin{table*}[tb]
\centering
\resizebox{1.8\columnwidth}{!}{%
\begin{tabular}{p{0.19\linewidth}  p{0.355\linewidth}  p{0.355\linewidth}}
\toprule
                                & \toolBF & \toolZThree \\
\midrule
Path generation                 & Bellman-Ford-Moore, Walk to the root; No acyclic paths & Pruning with heuristics; Any paths within the heuristics  \\
Path selection                  & Combines multiple sub-paths & Selects the highest revenue path \\
Manual DeFi modeling                 & Not required  & Required          \\
Captures non-cyclic strategies  & No            & Yes (e.g., bZx)   \\
Optimally chosen parameters     & No            & Yes (subject to inaccuracy of binary search) \\
Maximum Revenue                 & \MaxRevenueConfirmedBF         & \MaxRevenueConfirmedZThree           \\
Total Revenue (over \NumDays)                  & \TotalRevenueConfirmedBF      & \TotalRevenueConfirmedZThree          \\
Lines of code (Python)      & $300$           & $2,300$                 \\
\bottomrule
\end{tabular}
}
\caption{High-level comparison between \toolBF and \toolZThree.}
\label{tab:comparison}
\end{table*}

\subsection{Limitations}
\label{sec:limitation}
We elaborate on a few limitations of our work.

\point{State dependency}
In this study, we focus on block-level state dependencies (cf.\ Appendix~\ref{ref:state_dependency}), i.e., we consider a state to only change when a new block is mined. In practice, a DeFi state can change several times within the same blockchain block (as several transactions can trade on a DeFi platform within a block). Our assumption hence may cause us to not consider potentially profitable trades. An alternative approach to study state dependency, which we leave to future work, is to perform a transaction-level analysis. Such an analysis would assume that the trader observes the peer-to-peer network layer of the Ethereum network. Based on the information of transactions in the memory pool (the pool of unconfirmed transactions), the transaction order and state changes in the next block could be estimated ahead of the block being mined. 

\point{Scalability}
One problem of \tool is the combinatorial path explosion. To mitigate this problem, heuristics reduce the path space, which only needs to be executed once. For every new block, \tool can parallelize the parameter search process to find the most profitable paths. A limitation of negative cycle detection is that it has to search for negative cycles before starting to search parameters. The graph needs to be updated after executing every strategy. This is difficult to parallelize and limits the system's real-time capability, especially when there are multiple negative cycles, or the cycle length is long. 


\point{Manual Modeling and Code Complexity}
\toolBF only needs to be aware of the spot price of each market and treats the underlying smart contracts and exchange protocols as a black box while greedily exploring opportunities. \toolZThree, however, requires the manual translation of the objective function into an SMT problem. This requires to encode the state transitions into a group of predicates (cf.\ Appendix~\ref{app:encoding_example}). The modeling process not only increases the code complexity (cf.\ Table~\ref{tab:comparison}) but also causes inaccuracies in the found solutions and therefore requires a validation process through, e.g., concrete execution.

\point{Approximated Revenue}
To avoid double-counting revenue when a profitable path exists over multiple blocks, we apply a state dependency analysis and only exploit paths with a state change (cf.\ Section~\ref{ref:state_dependency}). However, \tool's reported revenue is not accurate because: (i) We work on historical blockchain states. In practice, the profitability of \tool will be affected by the underlying blockchain's network layer; (ii) For simplicity within this work, we assume that \tool does not change other market participants' behavior. In practice, other traders are likely to monitor our activity and adjust their trading strategy accordingly.

\point{Multiple Traders}
\label{sec:multiple}
Within this work, we only consider a single trader using \tool. Zhou~\etal~\cite{zhou2020high} simulated the outcome of competing transactions from several traders under a reactive counter-bidding strategy. We believe that those results translate over to MEV when multiple traders (specifically non-miners) compete over \tool transactions. Zhou~\etal~\cite{zhou2020high}'s results suggest that the total revenue will be divided among the competing traders.

\section{Experimental Evaluation}\label{sec:experimental_evaluation}
To query the Ethereum blockchain, we set up a full archive Geth\footnote{\url{https://github.com/ethereum/go-ethereum}} node  (i.e., a node which stores all intermediate blockchain state transitions) on a 
AMD Ryzen Threadripper $3990$~X Processor ($4.3$~GHz,~$64$ cores), $4$x$2$ TB NVMe SSD RAID $0$ and $256$~GB RAM.
We perform the concrete execution with a custom py-evm\footnote{\url{https://github.com/ethereum/py-evm}}, which can fork the Ethereum blockchain at any given block height. To simplify our experimental complexity, we do not consider trades which yield below~\Ether{0.1} and are aware that this potentially reduces the resulting financial gain.

We select~\NumActions actions from the Uniswap, Bancor, and MakerDAO, with a total of~\NumAssets assets (cf.\ Table~\ref{tab:assets} and~\ref{tab:actions} in Appendix). To enable action chaining, all considered assets trade on Uniswap and Bancor, while SAI and DAI are convertible on MakerDAO. The total value of assets on the three platforms sums up to~$3.3$ billion USD, which corresponds to~$82\%$ of the total USD value locked in DeFi as of May~2020.

Both \toolBF and \toolZThree apply dependency-based state reduction. Stationary blockchain states are identified and skipped to avoid redundant computation and double counting of revenue.


\subsection{\toolBF}
We translate the~\NumAssets assets and ~\NumActions actions into a graph with~\NumAssets nodes and~$94$ edges. Each node in the graph represents a cryptocurrency asset. For each edge $e_{i,j}$ pointing from asset $c_i$ to $c_j$, we find all markets with asset $c_i$ as input, and output asset $c_j$. Each edge's weight is derived using the highest price found among all supporting markets, or $0$ if there is no market. We then follow Algorithm~\ref{alg:bf} to greedily extract arbitrage revenue as soon as one negative cycle is found. We use the BFCF (Bellman-Ford-Moore, Walk to the root) algorithm to find negative cycles, which operates in $O(|N^2|\cdot|E|)$. For each arbitrage opportunity, \toolBF gradually increases the input parameter (amount of \emph{base} cryptocurrency asset) until the revenue ceases to increase.

\subsection{\toolZThree}\label{sec:complexity}
We translate DeFi states into Z3~\cite{de2008z3} as constraints on state symbolic variables (cf.\ Section~\ref{sec:defi_modeling}). We symbolically encode all variables using floats instead of integers because the EVM only supports integers. Most DeFi smart contracts express floats as integers by multiplying floats with a large factor. Division and power are, therefore, estimated using integer math. This practice may introduce a bias in our state and transition functions. Due to such model inaccuracies, we proceed to concrete execution (i.e., real-world smart contract execution on the EVM) to avoid false positives and validate our result.

An exhaustive search over the total action space is infeasible. Therefore, we apply path pruning (cf. Section~\ref{sec:pruning}) to discard irrelevant paths.


\point{Path Discovery and Pruning}
The~\NumActions DeFi actions (cf.\ Table~\ref{tab:actions} in Appendix) result in~$9.92 \times 10^{149}$ possible paths without repeating actions, which is an impractical space to evaluate. Table~\ref{tab:heuristics} hence illustrates the impact of our heuristics on paths of various lengths. We observe a significant reduction of at least~$99.98\%$ per path length of the total number of considered paths, resulting in only~\NumPaths remaining paths. The majority of paths~($77.67\%$) consist of~4 actions, while the shortest paths count~$2$ actions, and the longest~$5$ actions. Although we do not enforce a constraint on the maximum number of actions, all paths with more than~$5$ actions failed to pass our heuristics.

\begin{table}[tb]
\centering
\setlength{\tabcolsep}{5pt}
\begin{tabular}{lrrrrrrrr}
\toprule
\bf Path length & \bf Before & \bf After \\
\midrule
2     & $9,120$         & $2$   \\
3     & $857,280$       & $90$  \\
4     & $79,727,040$    & $466$ \\
5     & $7,334,887,680$ & $42$  \\
\midrule
\bf Total & $7,415,481,120$ & $600$ \\
\bottomrule
\end{tabular}
\caption{
Results of path pruning after applying the heuristics from Section~\ref{sec:pruning}. In total,~\NumPaths paths remain, with the majority~($77.67\%$) consisting of~4 actions. For each path length, the heuristics remove at least~$99.98\%$ of the strategies.}
\label{tab:heuristics}
\end{table}

\point{Action Dependency}
In the following, we present a concrete example of determining the dependency between two actions. The first action $a_{\text{Uniswap}}$ transacts ETH to SAI using the Uniswap SAI market. The second action $a_{\text{Bancor}}$ transacts BNT to SAI using the Bancor SAI contract. Equation~\ref{eq:storage1} and Equation~\ref{eq:storage2} show the relevant storage variables, respectively. These two actions are not independent, as they both modify the trader's balance in the SAI contract.
\begin{equation}
\begin{aligned}
\mathcal{K}^{\mathbb{T}}(a_\text{Uniswap}) = \{ &\text{<UniswapSAI>}.\text{ETH}, \\
&\text{<SAI>}.\text{balance of UniswapSAI}, \\
&\text{<Trader>}.\text{ETH}, \\
&\text{<SAI>}.\text{balance of trader} \} \\
\end{aligned}
\label{eq:storage1}
\end{equation}
\begin{equation}
\begin{aligned}
\mathcal{K}^{\mathbb{T}}(a_\text{Bancor}) = \{ &\text{<BNT>}.\text{balance of BancorSAI} \\
&\text{<SAI>}.\text{balance of BancorSAI} \\
&\text{<BNT>}.\text{balance of trader} \\
&\text{<SAI>}.\text{balance of trader} \} \\
\end{aligned}
\label{eq:storage2}
\end{equation}

\begin{figure}[tb]
\centering
\includegraphics[width=\columnwidth]{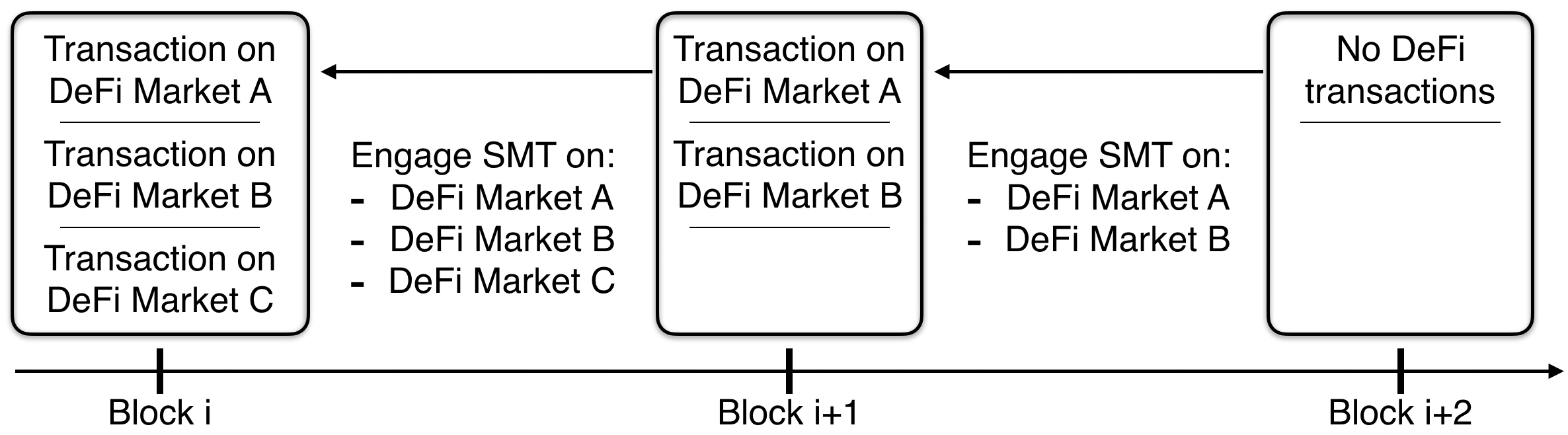}
\caption{We consider each blockchain block as an independent state representation of the DeFi platform markets. Only if a DeFi market changes in state, we need to re-engage the SMT solver for the affected paths only.}
\label{fig:block-dependency}
\end{figure}

\point{Dependency-based Blockchain State Reduction}
If a DeFi state does not change across a number of blockchain blocks, the same SMT solver computation is not re-engaged (cf.\ Figure~\ref{fig:block-dependency}). Algorithm~\ref{alg:dependency} specifies the algorithm we apply to automate the dependency-based blockchain state reduction. Figure~\ref{fig:asset_dependency} in the Appendix shows a timeline analysis of the state dependencies for all considered assets. We observe that ETH experiences the most state changes with over~\NumBlocks blocks~(\ETHStateChange), followed by DAI~(\DAIStateChange).

\begin{algorithm}[]
\DontPrintSemicolon
\SetAlgoLined
\SetKwProg{Fn}{Function}{ is}{end}
\KwIn{\;
$p = (a_1, a_2, \ldots) \in P$ $\gets$ Path $;\,$ $b$ $\gets$ Block number\;
}
\KwOut{Has a state change}
\ForEach{$a \in p$}{
    \BlankLine
    \ForEach{$s \in \mathcal{K}^{\mathbb{T}}(a)$}{
        \If{fetch(\textit{s}, \textit{b}) $\neq$ fetch(\textit{s}, \textit{b - 1})}{
            \Return{\text{True}}\;
        }
    }
    \BlankLine
}
\Return{\text{False}}\;
\;
\Fn{fetch(\textit{s}, \textit{b})}{
    return (\textit{Concrete value for storage variable address $s$ on block $b$}) \;
}
\caption{Block state dependency analysis.}
\label{alg:dependency}
\end{algorithm}

\subsection{\toolBF and \toolZThree: Revenue}
\begin{figure}[tb]
\begin{center}
\includegraphics[width = 0.95\columnwidth]{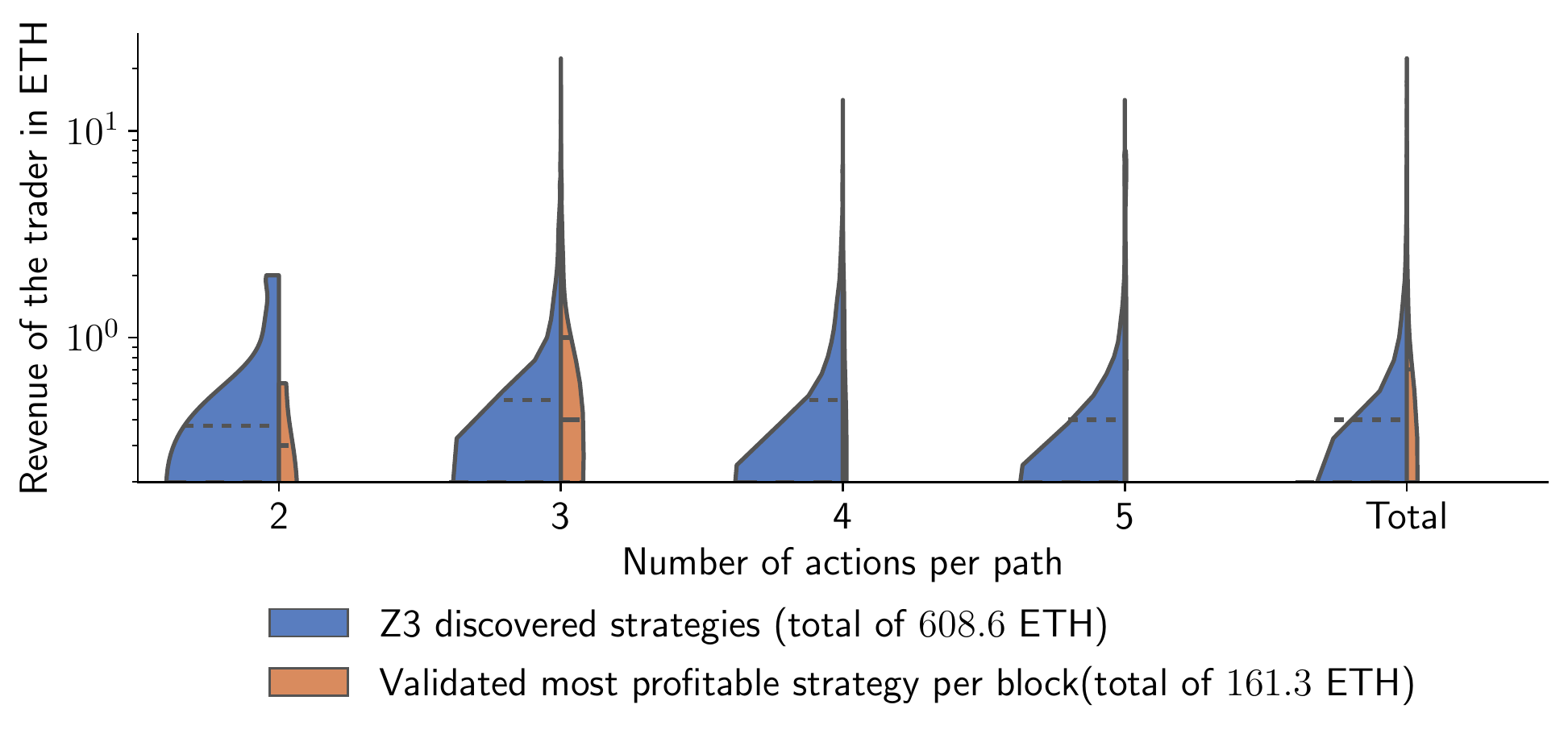}
\end{center}
\caption{Analytical distribution of the trader's revenue. The majority of the most profitable strategies consist of~3 actions. Because \toolZThree requires manual modeling, the revenues discovered by Z3 are not accurate, and thus the number of discovered strategies (yellow) are less than the profitable strategies (blue). We use concrete execution to validate the strategies from Z3.}
\label{fig:analytical_revenue}
\end{figure}

\begin{figure}[tb]
\begin{center}
\includegraphics[width = 0.95\columnwidth]{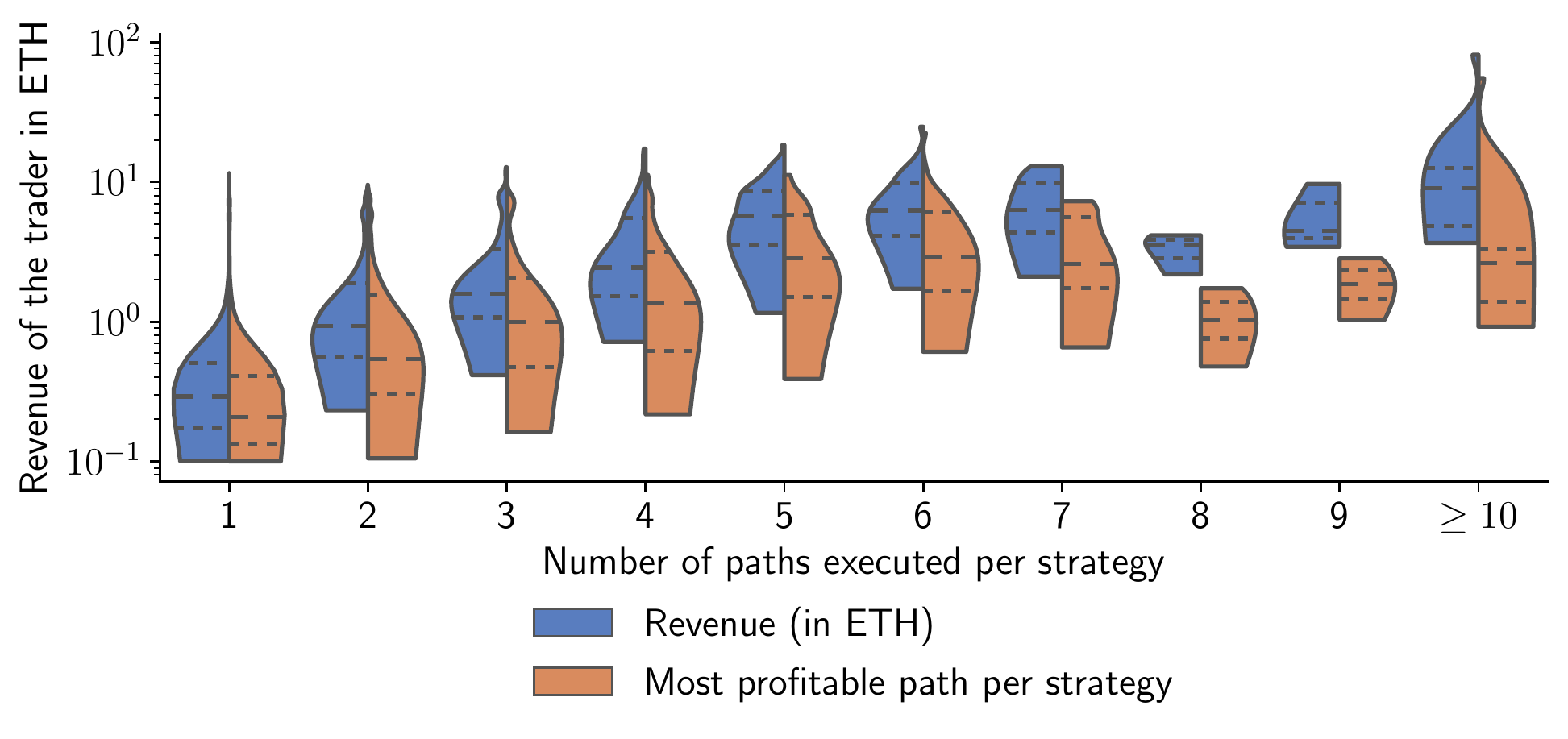}
\end{center}
\caption{Distribution of the trader's revenue using \toolBF. We observe that the revenue increases as the number of paths increases. We also visualize the distribution of the most profitable sub-path (orange) for every strategy. Intuitively, the more paths \toolBF try to combine, the higher the revenue.}
\label{fig:revenue_bf}
\end{figure}

We validate both \tool designs on past blockchain data from block~\StartingBlock to block~\EndingBlock, over a total of \NumDays. We visualize the distribution of traders' revenue for \toolZThree in Figure~\ref{fig:analytical_revenue}. \toolZThree found~\ZFoundStrategies strategies consisted of~$2$ to~$5$ actions. In total~\toolZThree yields a total of~\TotalRevenueConfirmedZThree, and we observe that the most profitable strategies consist of~$3$ actions, where the highest revenue yielded amounts to~\MaxRevenueConfirmedZThree. Similarly, Figure~\ref{fig:revenue_bf} visualizes the distribution of traders' revenue for \toolBF. Recall that \toolBF greedily combine multiple paths into a single strategy. We observe that the revenue increases as the number of paths increases, with the highest revenue amounting to~\MaxRevenueConfirmedBF. In total, \toolBF finds~\BFFoundStrategies strategies and yields~\TotalRevenueConfirmedBF.


\begin{figure*}[tb]
\centering
\begin{subfigure}{\columnwidth}
\includegraphics[width = 0.9\columnwidth]{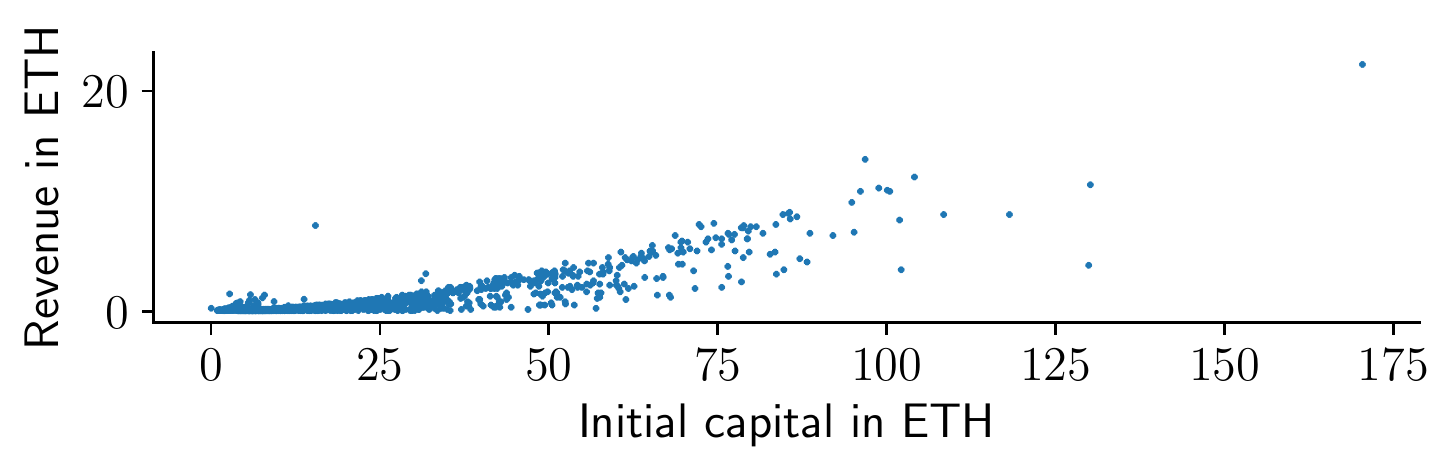}
\caption{\toolZThree without flash loans.}
\end{subfigure}
\hfill
\begin{subfigure}{\columnwidth}
\includegraphics[width = 0.9\columnwidth]{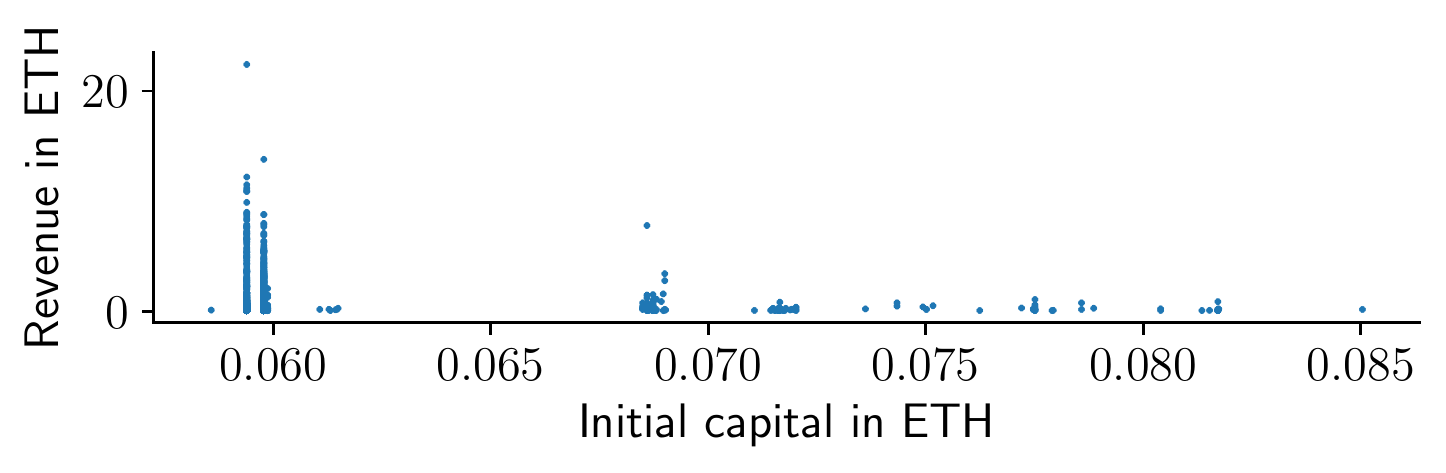}
\caption{\toolZThree with flash loans.}
\end{subfigure}
\vskip\baselineskip
\begin{subfigure}{\columnwidth}
\includegraphics[width = 0.9\columnwidth]{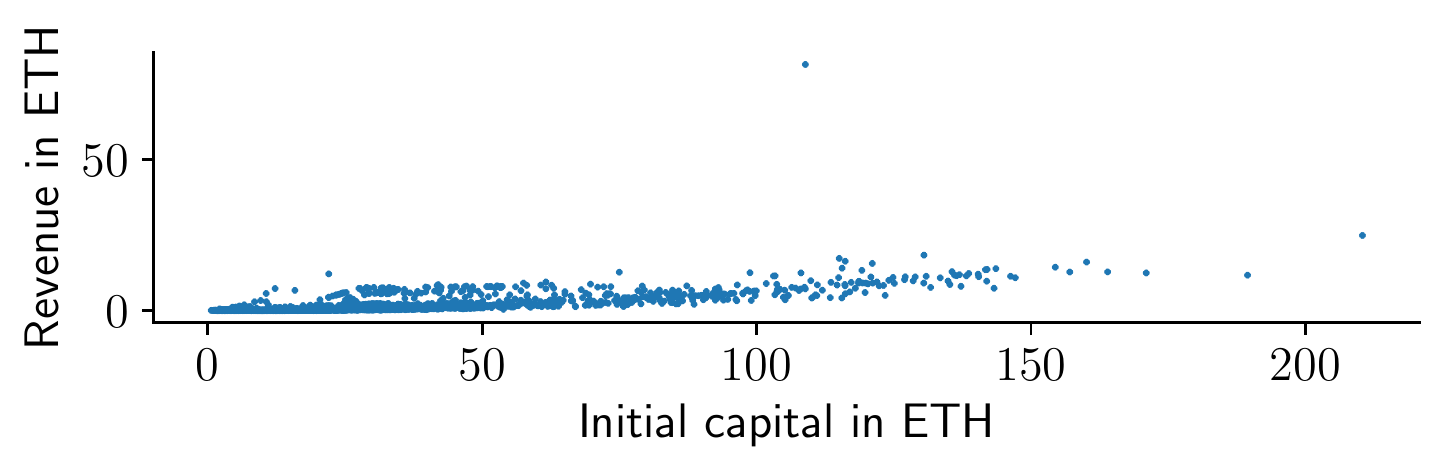}
\caption{\toolBF without flash loans.}
\end{subfigure}
\hfill
\begin{subfigure}{\columnwidth}
\includegraphics[width = 0.9\columnwidth]{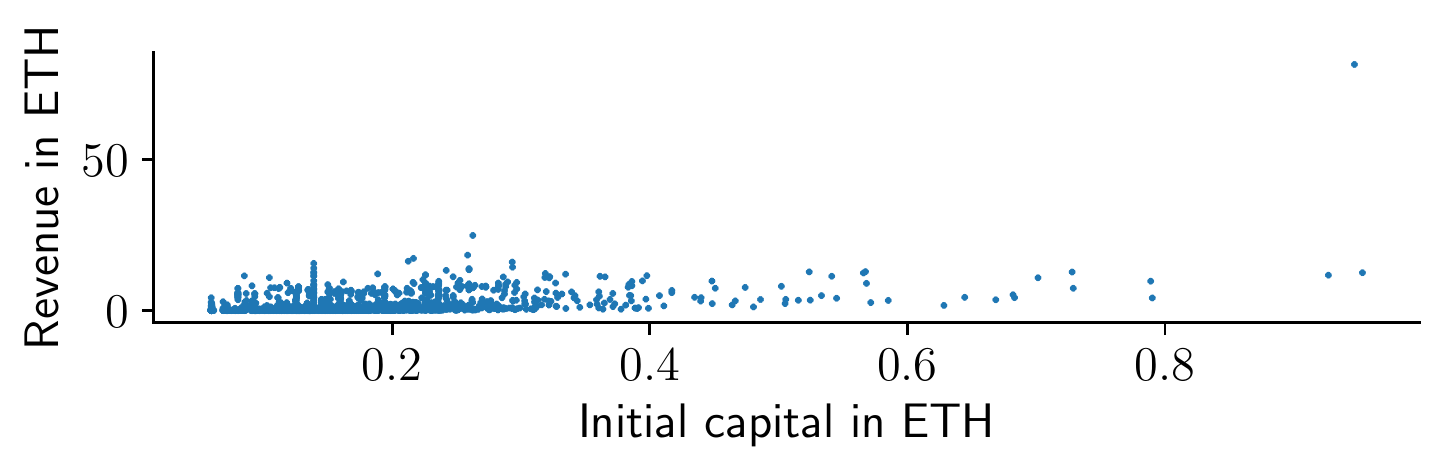}
\caption{\toolBF with flash loans.}
\end{subfigure}

\caption{Revenue as a function of the initial capital, in ETH with and without flash loans for \toolBF (total of ~\NumConfirmedStrategiesBF found strategies) and \toolZThree (total of~\NumConfirmedStrategiesZThree found strategies).}
\label{fig:revenue_as_a_function_of_input}
\end{figure*}

We visualize in Figure~\ref{fig:revenue_as_a_function_of_input} the revenue generated by \toolZThree and \toolBF as a function of the initial capital. If a trader owns the \emph{base} asset (e.g., ETH), most strategies require less than $150$~ETH. Only $10$ strategies require more than $100$~ETH for \toolZThree, and only $7$ strategies require more than $150$ ETH for \toolBF. This capital requirement is reduced to less than \Ether{1} when using flash loans (cf.\ Figure~\ref{fig:revenue_as_a_function_of_input} (b, d)). 

\begin{figure*}[tb]
\begin{subfigure}{\textwidth}
\centering
\includegraphics[width = 0.95\textwidth]{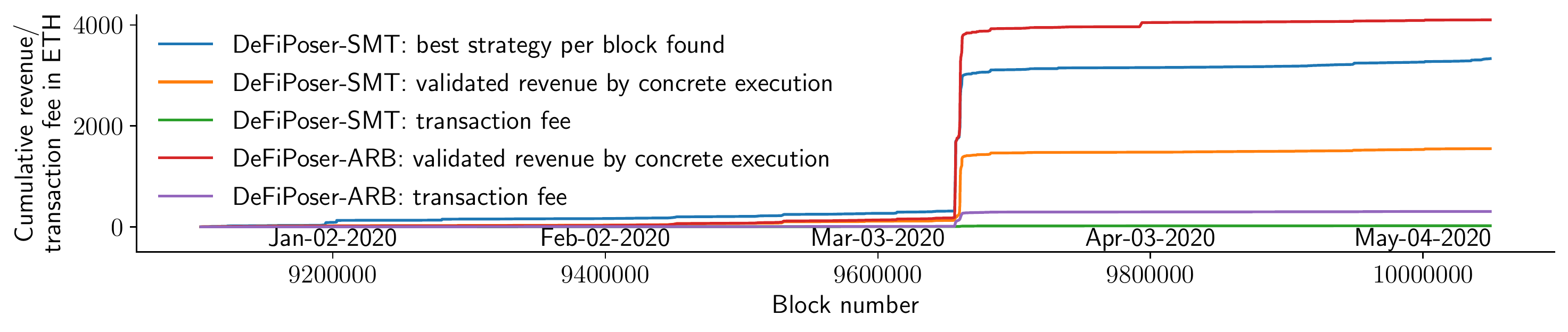}
\caption{Cumulative revenue over time (\NumDays), found by \toolZThree and \toolBF, and validated via concrete execution.}
\label{fig:revenue_over_time}
\end{subfigure}
\vskip\baselineskip
\begin{subfigure}{0.95\columnwidth}
\centering
\includegraphics[width = \textwidth]{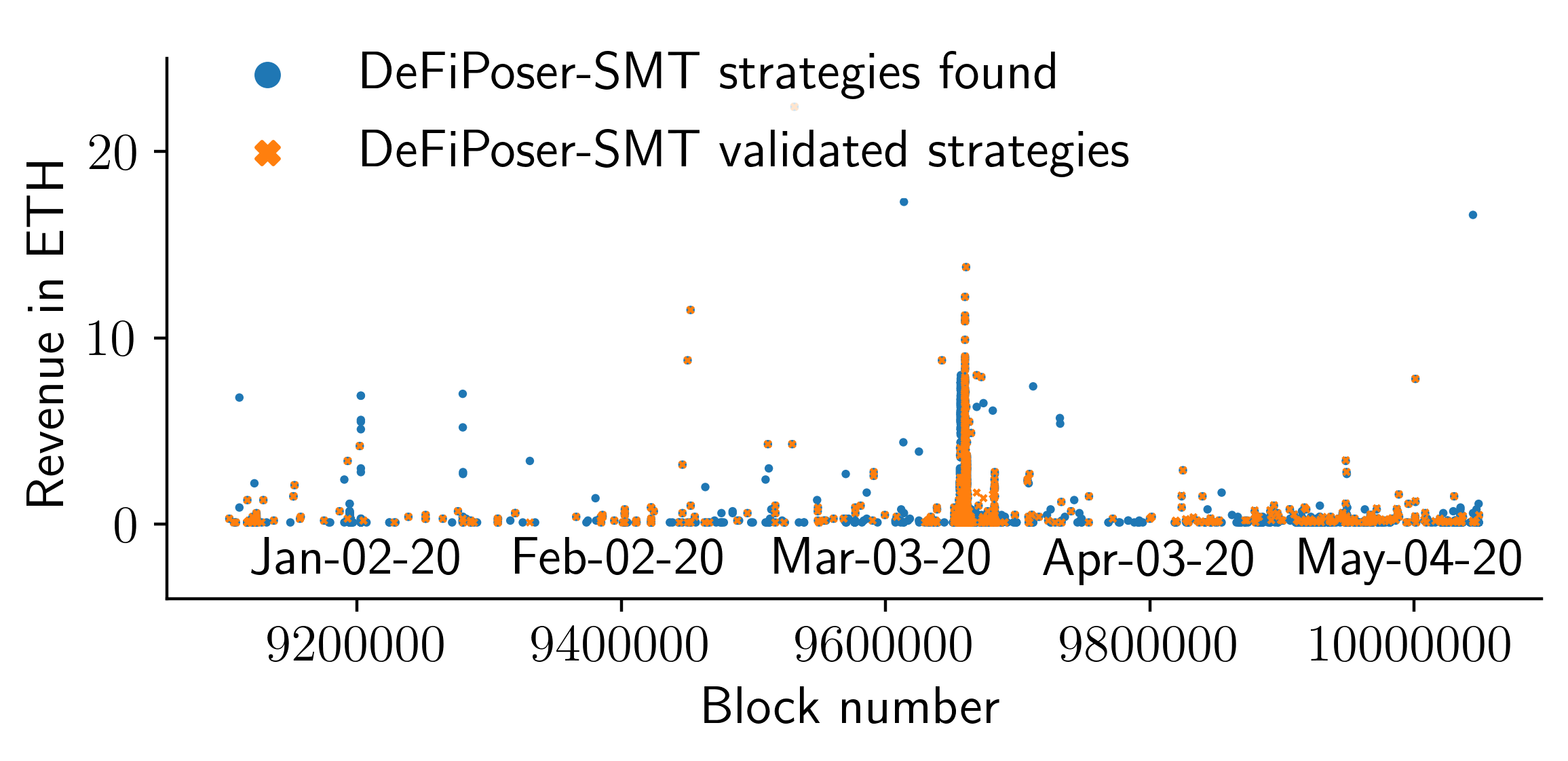}
\caption{Strategies detected and validated by \toolZThree.}
\label{fig:time-revenue}
\end{subfigure}
\hfill
\begin{subfigure}{0.95\columnwidth}
\centering
\includegraphics[width = \textwidth]{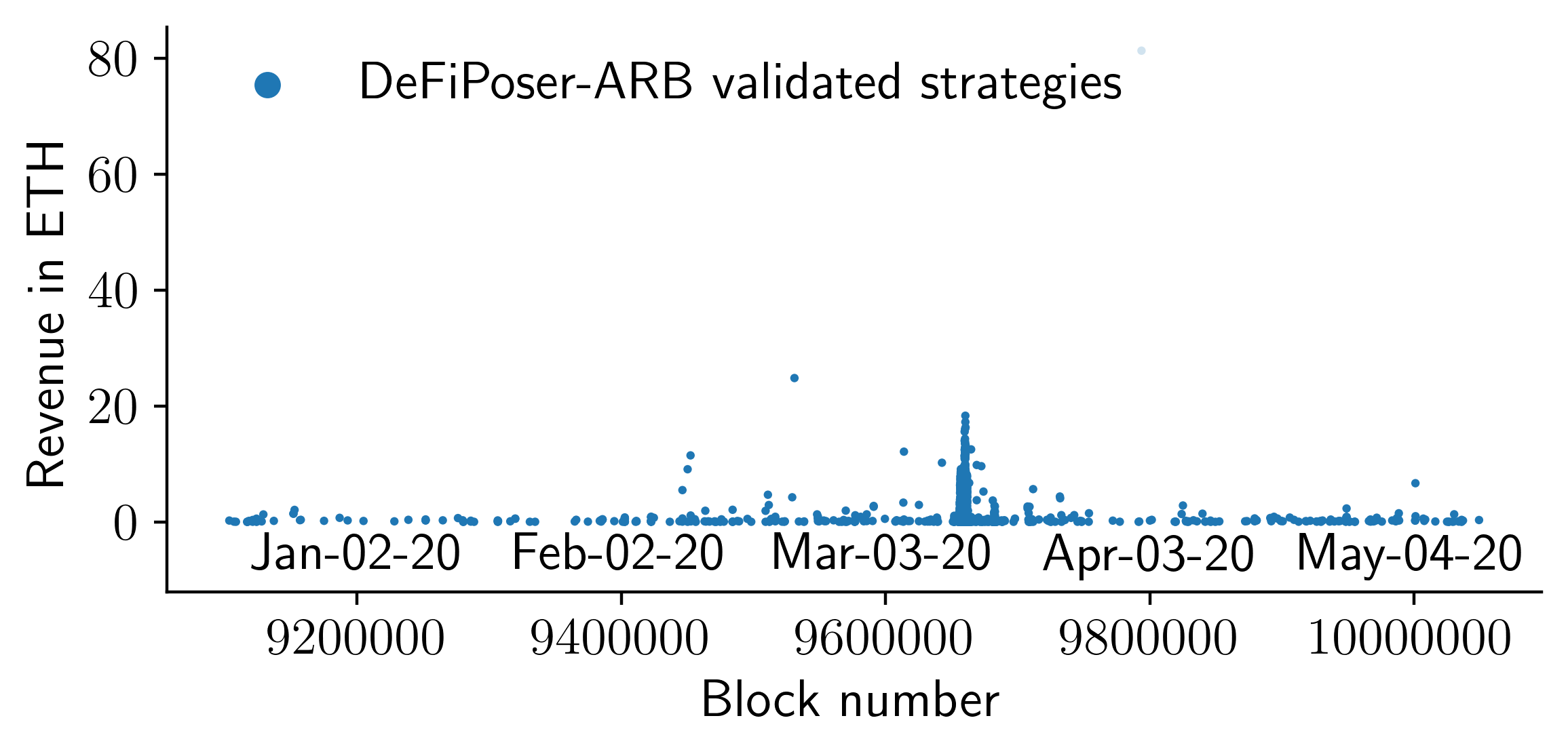}
\caption{Strategies validated by \toolBF.}
\label{fig:time-revenue}
\end{subfigure}

\caption{Revenue and transaction fees analysis over time, measured in blocks.}
\end{figure*}

Figure~\ref{fig:revenue_over_time} shows how our concrete execution validation over~\NumDays yields consistent revenue for both tools. The concrete execution estimates a weekly revenue of \WeeklyRevenueBF for \toolBF and \WeeklyRevenueZThree for \toolZThree. For \toolZThree, our validation estimates a total revenue of~\TotalRevenueConfirmedZThree out of~\TotalRevenueUnconfirmedZThree (i.e., 40\% of the Z3 indicated revenue is validated in practice).

\begin{figure}[tb]
\centering
\begin{subfigure}{0.95\columnwidth}
\includegraphics[width = \columnwidth]{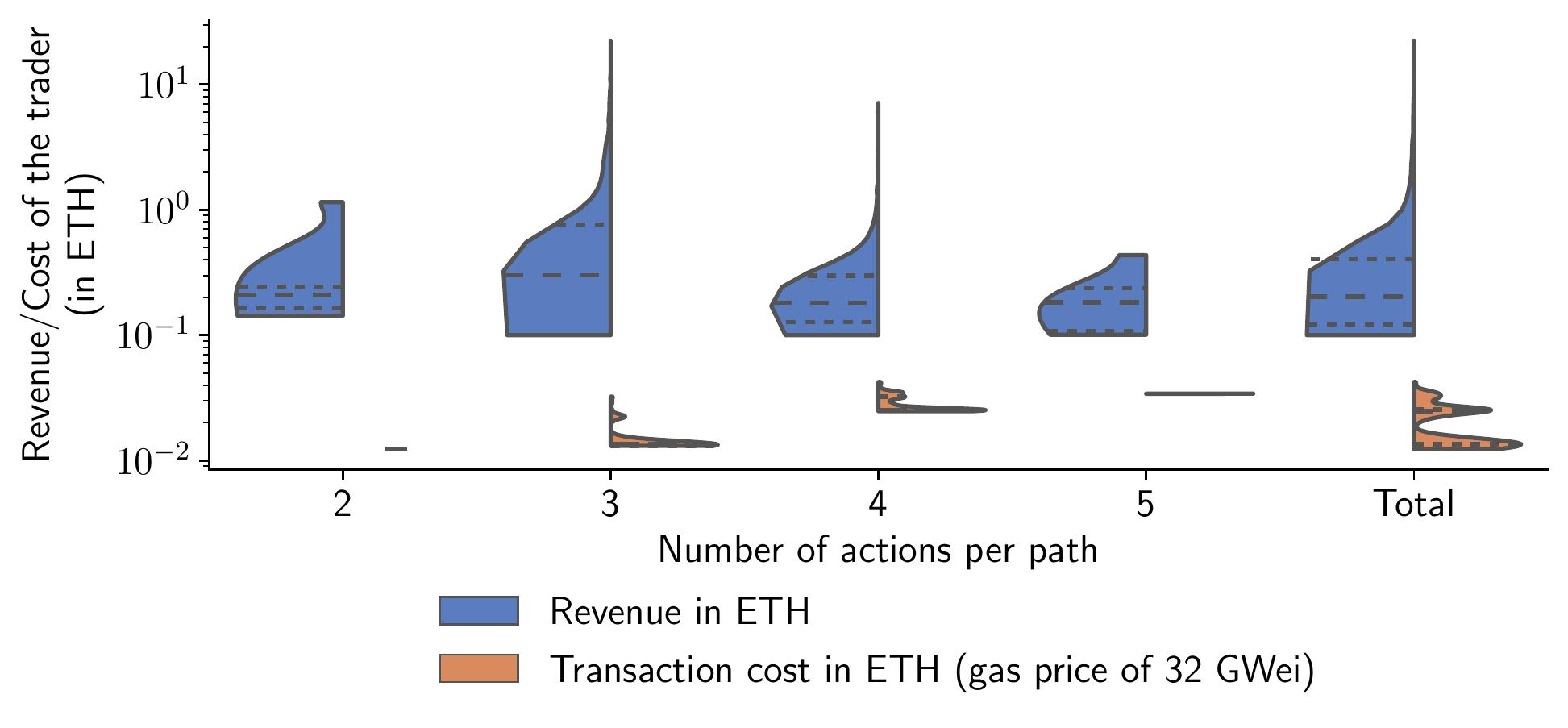}
\caption{\toolZThree}
\end{subfigure}
\begin{subfigure}{0.95\columnwidth}
\includegraphics[width = \columnwidth]{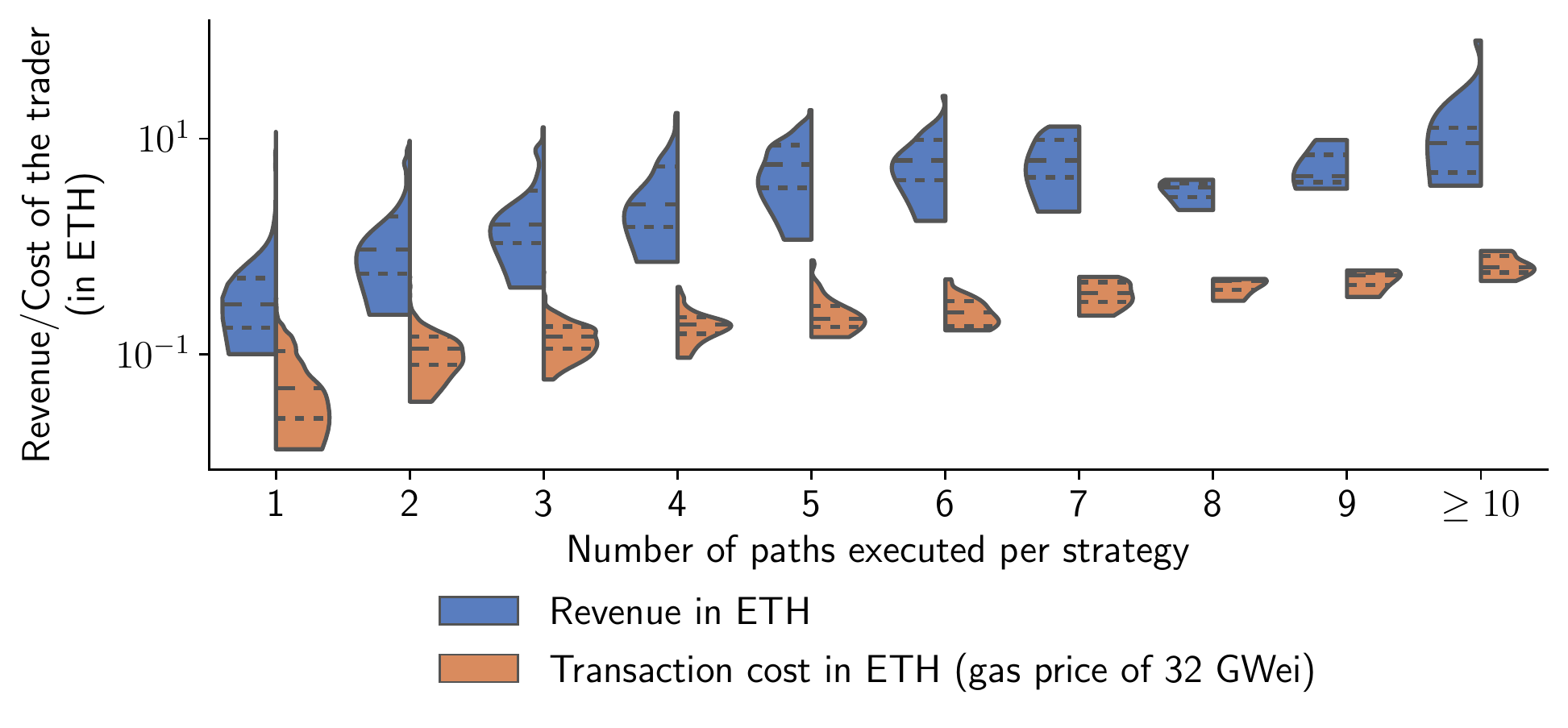}
\caption{\toolBF}
\end{subfigure}
\caption{Distribution of revenue and transaction cost based on concrete execution on the EVM for \toolZThree and \toolBF. The revenue outpaces the transaction costs, which are higher for \toolBF because the found strategies often consist of more cycles (arbitrage opportunities).}
\label{fig:validation_revenue}
\end{figure}

\point{Cost Analysis}
The trader's principal costs are the blockchain transaction fees (e.g., gas in Ethereum), which remain below the revenue yielded by the strategies we validated (cf.\ Figure~\ref{fig:validation_revenue}). Note that a trading strategy may fail if the underlying market state changes before its execution. Therefore, we assume that the trader adopts the gas price of~\FastGasPrice, which is highly volatile, but the recommended fast transaction gas price at the time of writing. Summarizing, the execution of all strategies costs less than~$0.05$~ETH, which warrants all strategies to be profitable.

\begin{figure}[tb]
\centering
\includegraphics[width=\columnwidth]{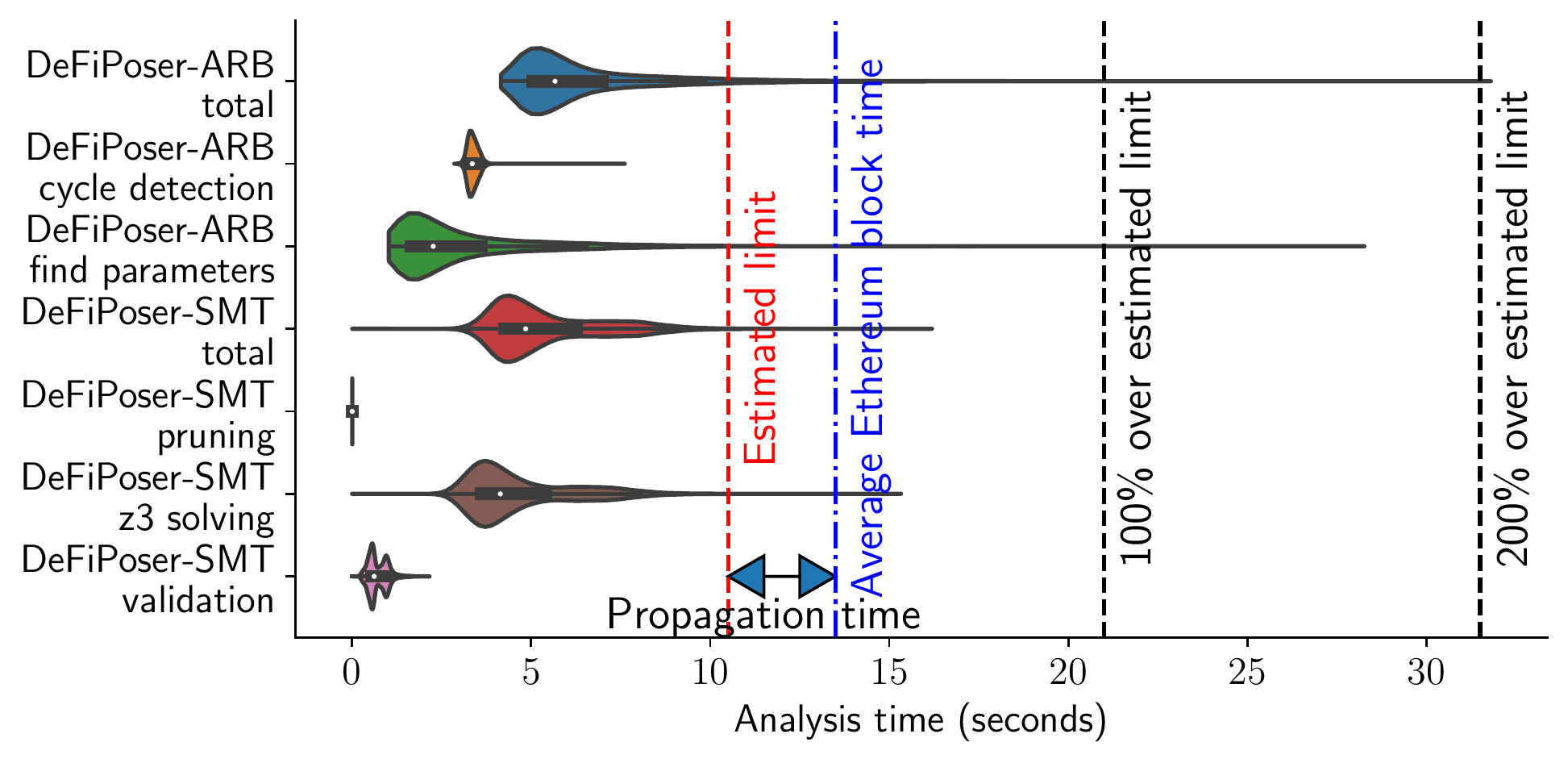}
\caption{Analysis time distribution to detect a profitable strategy on \toolZThree and \toolBF. For most strategies the search and validation process remains below the average Ethereum block time of~$13.5 \pm 0.12$ seconds.}
\label{fig:outliers}
\end{figure}

\begin{figure}[tb]
\centering
\includegraphics[width=\columnwidth]{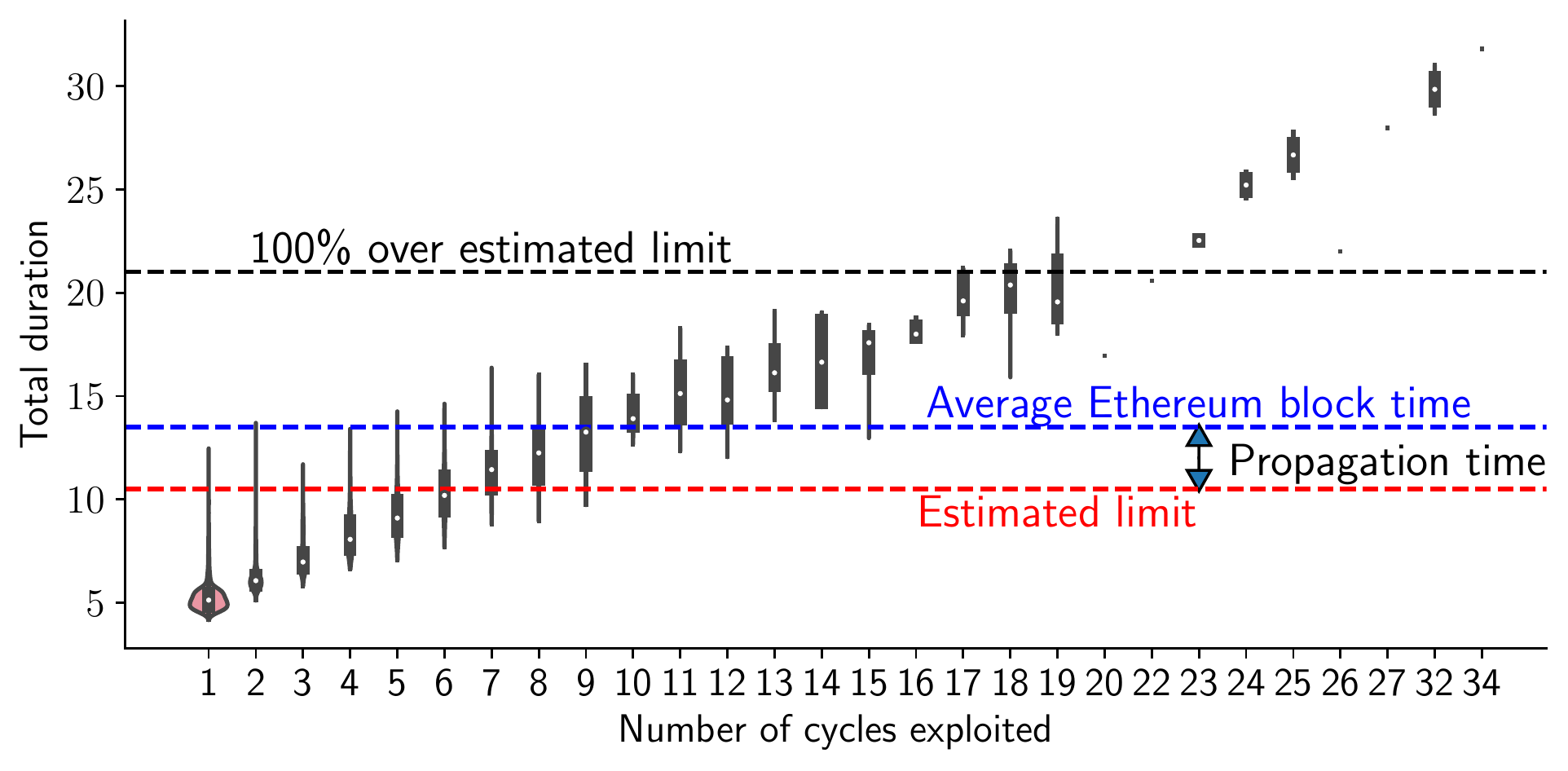}
\caption{The analysis time of \toolBF exceeds our estimated time limit (taking into consideration block time and transaction network propagation to miners), when \toolBF exploits more than~$6$ cycles.}
\label{fig:bf_timeout}
\end{figure}

\point{Performance Analysis} Our tools must find trades within the average Ethereum block time of $13.5 \pm 0.12$ seconds~\cite{ethereum-block-time} to be applicable in real-time. Assuming a network propagation latency of roughly three seconds towards miners in the blockchain P2P network~\cite{decker2013information}, our tools must generate transactions within at most $10.5$ seconds. Figure~\ref{fig:outliers} shows the detailed execution speed of \toolBF and \toolZThree on an AMD Ryzen Threadripper $3990$~X Processor ($4.3$~GHz,~$64$ cores) CPU. For new block states, we measure a total average computing time of \AverageAnalsysTimePerBlockBF and \AverageAnalsysTimePerBlockZThree per block, respectively.

We further group the strategies detected by \toolBF based on the number of negative cycles and compare the respective analysis time (cf.\ Figure~\ref{fig:bf_timeout}). We find that \toolBF exceeds our estimated time limit ($13.5 - 3 = 10.5$ seconds) when exploiting more than~$6$ cycles. The higher the total number of negative cycles, the more likely \toolBF misses the most profitable opportunity.

\section{Profitable Transactions and Blockchain Security}
\label{sec:security}
In this section, we show that \toolZThree is capable of identifying the economic bZx attack from February 2020~\cite{qin2020attacking} and provide forensic insights into the event. Given optimal adversarial strategies provided by an MDP, we then quantify whether an MEV opportunity will cause a rational miner to create a blockchain fork.

\subsection{Economic bZx Attack}
On the~15th of February,~2020, a trader performed a pump and arbitrage attack on the margin trading platform bZx\footnote{transaction id: \wrapletters{0xb5c8bd9430b6cc87a0e2fe110ece6bf527fa4f170a4bc8cd032f768fc5219838}}. The core of this trade was a pump and arbitrage involving four DeFi platforms atomically executed in one single transaction. As a previous study shows, this trade resulted in in~\Ether{4337.62} loss from bZx loan providers, where the trader gained~\Ether{1193.69} in total~\cite{qin2020attacking}.

\point{Attack Window}
To gain deeper insights into this DeFi composability event, we extend \toolZThree with two additional actions: \emph{(i)} borrow WBTC with ETH on compound finance; \emph{(ii)} short ETH for WBTC on bZx. We replayed \toolZThree on historical blockchain data by starting at the creation of the bZx's margin short smart contract (cf.\ Figure~\ref{fig:bzx_attack_window}). Surprisingly, the bZx attack window lasted for~$69$ days until it was openly exploited. \toolZThree finds that the attack yielded the highest revenue of \Ether{2291.02} at block~$9,482,670$, which is about one day before the attack occurred.

\subsection{MEV, an MDP and Optimal Adversarial Strategies}
\label{sec:mdp}

The economic bZx attack revenue exceeds the average Ethereum block reward\footnote{At the time of writing, the average Ethereum block reward is $2.62$ ETH (\url{https://bitinfocharts.com/ethereum/})} by a factor of \BZXExceedsBlockReward. After bZx, the other most profitable validated strategies found by \toolBF and \toolZThree exceed the block reward by a factor of \BFExceedsBlockReward and \ZTHREEExceedsBlockReward respectively. In this section, we quantify the value at which an MEV-aware miner would exploit an MEV opportunity by forking the blockchain.

\point{Markov Decision Process}
A Markov Decision Process is a single-player decision process that allows identifying the optimal strategies for an encoded decision problem. In this work, we adopt the state transition and reward matrix of the PoW double-spending MDP of Gervais \emph{et al.}~\cite{gervais2016security}. Note that the MDP we use does not consider uncle rewards.

We observe that conceptually, an MEV opportunity is equivalent to a double-spending opportunity: if an MEV opportunity is mined by an honest miner, and an adversarial miner aims to claim the MEV opportunity, the MEV miner will need to outrun the honest chain with a fork. The MEV miner will hence want to follow the optimal adversarial strategies given by the MDP, which advises whether to fork or not to fork the blockchain depending on the MEV value.

\point{Threat Model}
We assume a rational and computationally bounded adversary. Because MDP's are single-player decision problems, we assume the existence of only one adversarial miner willing to exploit MEV. We parametrize the miner with a hash rate $\alpha \in ]0,0.5[ $, while the remaining non-MEV miners have a hash rate of $1-\alpha$. We ignore the existence of eclipse attacks ($\omega = 0$) and assume the weakest possible network propagation parameter of the adversary ($\gamma = 0$). We parametrize the MDP with the stale block rate of the Ethereum blockchain at the time of writing. By crawling the number of uncle blocks (from the block $9.1$M to $10.5$M), we approximate the stale block rate to $r_s=5.72\%$. We set the mining costs to match the hash rate of the MEV miner ($c_m = \alpha$).

\noindent\textbf{Exploit or not exploit MEV?}
Each time an MEV opportunity arises on the network layer, we assume that the honest miner succeeds in mining the MEV opportunity, and the MEV miner fails to receive the reward initially. The MEV miner, therefore, needs to decide whether to start to mine on a private chain, where he claims the MEV opportunity. Note that the MDP's $exit$ state can only be reached when the MEV miner mined a private chain that is longer than the honest chain ($l_a > l_h$) given $k=1$ ($l_a > k$). \emph{Depending on the MEV value, the optimal strategy $\pi$ might advise against forking the chain to attempt to claim the MEV reward.} We quantify the minimal MEV value $MEV_v$, such that $MEV_v$ is strictly larger than the reward from honest mining (cf.\ Equation~\ref{eq:mev-v}). We denote $h$ is the process of mining honestly.
\begin{equation}\label{eq:p}
P = (\alpha, \gamma, r_s, k, \omega, c_m)
\end{equation}
\begin{equation}\label{eq:mev-v}
    MEV_v = min\{MEV_v | \exists \pi \in A: R(\pi,P,MEV_v) > R(h, P)\}
\end{equation}

To solve the MEV MDP for the optimal strategies, we use the code of~\cite{gervais2016security}\footnote{\url{https://github.com/arthurgervais/pow_mdp}} and reparametrize given the current Ethereum stale block rate ($r_s=5.72\%$). We further set $k=1$, $\gamma=0$, $\omega=0$ and the cut-off value (the maximum length of $l_a$ and $l_h$) to $20$ blocks. Similar to~\cite{gervais2016security}, we apply a binary search to find the lowest value for $MEV_v$ in units of block rewards, given a margin of error of $0.1$.

\begin{figure}[tb]
\centering
\includegraphics[width=\columnwidth]{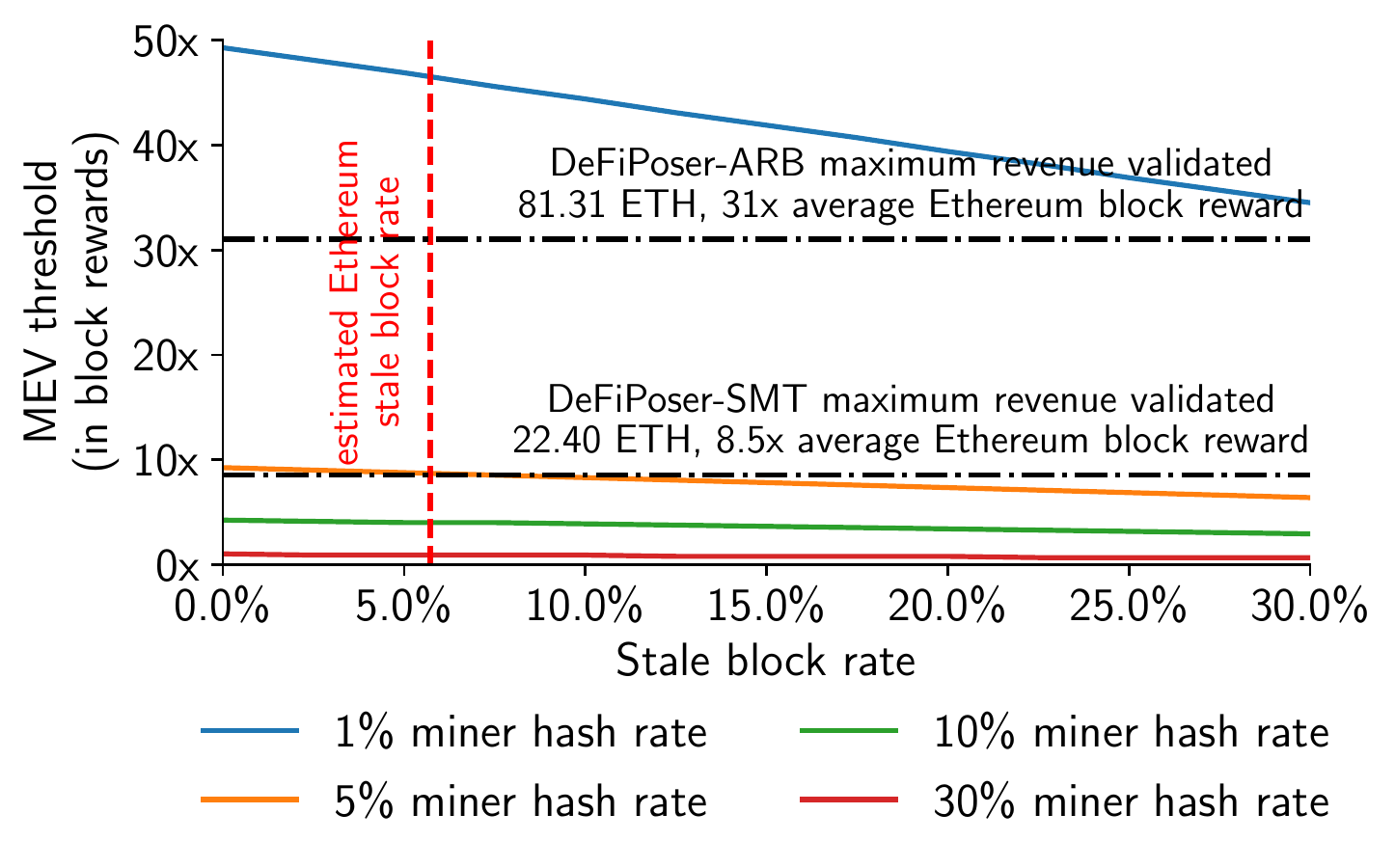}
\caption{Minimum MEV value in terms of block rewards to fork a PoW blockchain, given by optimal adversarial strategies of the MDP. For instance, on Ethereum ($r_s=5.72\%$), a miner with $10$\% hash rate will engage to fork the chain to exploit an MEV opportunity, if the adversary follows the optimal strategy and the MEV opportunity yields more than $4$ block rewards.}
\label{fig:mev-v}
\end{figure}

\point{Results} We visualize our findings in Figure~\ref{fig:mev-v}, which shows that for an MEV miner with $10$\% hash rate, on Ethereum (stale block rate of $5.72$\%), $MEV_v$ equals to $4$. We conclude that in this case, if an MEV opportunity yields at least a reward that is $3$ times higher than the block reward, then an MEV miner which follows the optimal strategies will fork the blockchain. A fork of the blockchain deteriorates the blockchain's security as it increases the risks of double-spending and selfish mining~\cite{gervais2016security}.

\point{Multiple MEV Miner}
Our MDP model does not allow us to draw conclusions on the dynamics under multiple independent MEV miners. We hence can only speculate about the outcome and leave a simulation to future work. We can imagine that multiple miners either collaborate to share an MEV profit (which falls back to our MDP game of one adversary), or the miners would compete among each other, which is likely to exacerbate the fork rate and hence further deteriorates the blockchain consensus security.

\section{Related Work}
\label{sec:relatedwork}

While the research literature of blockchain span over~10 years, DeFi is a relatively recent area with fewer works.

\point{DeFi}
There is a growing body of literature focusing on the security of the DeFi ecosystem. Blockchain front-running in exchanges, games, gambling, mixer, the network layer, and name services is soundly studied~\cite{eskandari2019sok, daian2019flash, breidenbach2018enter, kalodner2015empirical, le2020amr, gervais2015tampering, ConsenSy8:online, zhou2020high}. Daian~\etal~\cite{daian2019flash} demonstrate a thorough analysis of profiting from opportunities provided by transaction ordering.
Xu~\etal~\cite{livshits19pump-and-dump} presents a detailed study of a specific market manipulation scheme, pump-and-dump, and build a prediction model that estimates the pump likelihood of each coin.
Gudgeon~\etal~\cite{gudgeon2020decentralized} explore the possibility of a DeFi crisis due to the design weakness of DeFi protocols and present a stress testing framework.
Qin~\etal~\cite{qin2020attacking} investigate DeFi attacks through flash loans and how to optimize their profit. We remark that the optimization solution presented in~\cite{qin2020attacking} only applies to previously fixed attack vectors, while this work considers the composability of DeFi protocols.

\point{Smart Contract Analysis}
Besides the above-mentioned works on DeFi, many studies on the vulnerability discovery of smart contracts are related to our work~\cite{tsankov2018securify, brent2018vandal, grech2018madmax, luu2016making, krupp2018teether, chang2019scompile, kalra2018zeus, tikhomirov2018smartcheck, jiang2018contractfuzzer, echidna2020, Harvey2016, bohme2017directed, he2019ilf}. Traditional smart contract vulnerabilities examined in related work include, for instance, re-entrancy attack, unhandled exceptions, locked ether, overflow~\cite{luu2016making}. To the best of our knowledge, no analysis tool has yet considered the problem of a \emph{composability analysis} as we've performed.

\point{Model Checking}
Model-checking is another viable method to verify the security of smart contracts. Model-checking examines all possible states in a brute-force manner~\cite{baier2008principles} and performs systematic exhaustive exploration for checking whether a finite transition machine model of a system meets appropriate specifications~\cite{tsankov2018securify, brent2018vandal, grech2018madmax, luu2016making, krupp2018teether, chang2019scompile, kalra2018zeus, tikhomirov2018smartcheck}. One of the main limitations of model-checking is the exponential growth of the number of possible states, resulting in unsolvability for complex contracts.

\section{Conclusions}
\label{sec:conclusion}
This paper presents two practical approaches that automatically extract revenue from the intertwined mesh of decentralized finance protocols. The first technique, \toolBF, is well-suited for arbitrage, and the second, \toolZThree, can also find acyclic opportunities. When evaluated over a span of~\NumDays with~\NumActions DeFi actions and~\NumAssets cryptocurrency assets, \toolBF and \toolZThree are estimated to generate an average weekly revenue of~\WeeklyRevenueBF and~\WeeklyRevenueZThree, with the highest transaction being~\MaxRevenueConfirmedBF and~\MaxRevenueConfirmedZThree,  respectively.

Our techniques apply to a real-time operation on blockchains with reasonably fast inter-block times (such as Ethereum), with an average search of~\AverageAnalsysTimePerBlockBF and~\AverageAnalsysTimePerBlockZThree per block for \toolBF and \toolZThree, respectively, using a relatively unoptimized implementation. We find that the capital requirements to extract the found revenues are minimal: the majority of strategies produced require less than~\CapitalRequrementWithoutFlashLoan, without, and less than~\Ether{1} with flash loans.

We quantitatively demonstrate some troubling security implications of profitable transactions on the blockchain consensus. Given optimal adversarial strategies provided by a Markov Decision Process, we quantify the threshold value at which an MEV-aware rational miner will fork the blockchain if the miner does not succeed in claiming an unconfirmed MEV opportunity first. For example, on the current Ethereum network, a~$10$\% hash rate miner will fork the chain if an MEV opportunity exceeds~$4$ block rewards. As a comparison, the bZx opportunity exceeded the Ethereum block reward by a factor of~\BZXExceedsBlockReward! Our work hence quantifies the inherent tension between revenue extraction from profitable transactions and blockchain security. We can generally expect trading opportunities highlighted in this paper to expand as the DeFi ecosystem grows and becomes more popular.

\section*{Acknowledgments}
We very much thank the anonymous reviewers and Nicolas Christin for the thorough reviews and helpful suggestions that significantly strengthened this paper. We are moreover grateful to the Lucerne University of Applied Sciences and Arts for generously supporting Kaihua Qin's Ph.D.

\clearpage

\bibliographystyle{plain}
\bibliography{references.bib}

\appendices

\section{Summary of the ERC-20 cryptocurrency assets}

We summarize the \NumERC ERC-20 cryptocurrency assets in Table~\ref{tab:assets}. We observe that for most of the assets, the number of holders and the number of markets increases with the number of transfer transactions.

\begin{table}[tb]
\centering
\begin{tabular}
{>{\raggedleft\arraybackslash}p{1cm}%
   >{\raggedleft\arraybackslash}p{1.5cm}%
   >{\raggedleft\arraybackslash}p{2.5cm}%
   >{\raggedleft\arraybackslash}p{1.5cm}%
}
\toprule
\bf  Token        & \bf Unique holders & \bf Transfer transactions  & \bf  Markets trading 
\\
\midrule
SAI   & 181,223 & 3,139,071 & 4 \\ 
BNT   & 23,966  & 2,620,652 & 144 \\ 
DAI   & 68,357  & 2,155,535 & 130 \\ 
BAT   & 288,970 & 1,970,176 & 218 \\
ENJ   & 52,341  & 902,471   & 66 \\ 
SNT   & 82,663  & 868,007   & 101 \\
KNC   & 65,018  & 820,501   & 73 \\ 
MKR   & 20,891  & 733,845   & 67 \\ 
DATA  & 444,833 & 588,097   & 26 \\ 
MANA  & 38,276  & 565,151   & 77 \\ 
ANT   & 22,321  & 217,657   & 24 \\
RLC   & 12,880  & 209,255   & 24 \\ 
RCN   & 19,831  & 203,893   & 24 \\
UBT   & 10,410  & 191,153   & 14 \\
GNO   & 10,695  & 170,507   & 21 \\ 
RDN   & 13,842  & 143,308   & 16 \\
TKN   & 5,485   & 84,912    & 7 \\ 
TRST  & 7,738   & 71,223    & 7 \\
AMN   & 2,593   & 53,010    & 3 \\
FXC   & 2,024   & 47,906    & 14 \\
SAN   & 2,247   & 36,054    & 7 \\ 
AMPL  & 1,931   & 31,124    & 10 \\
HEDG  & 1,709   & 30,770    & 17 \\
POA20 & 560     & 26,390    & 10 \\
\bottomrule
\end{tabular}
\caption{
Summary of the \NumERC ERC-20 cryptocurrency assets used in our experiments, ordered by the total number of transfer transactions.}
\label{tab:assets}
\end{table}

\section{Supported DeFi actions}

We summarize the \NumActions DeFi actions \tool supports in Table~\ref{tab:actions}. All considered cryptocurrency assets trade on both the Uniswap and Bancor exchanges. SAI and DAI, in addition, can be converted to each on MakerDAO.

\begin{table}[tb]
\centering
\begin{tabular}{cccccc}
\toprule
\multicolumn{2}{c}{\textbf{Uniswap}} & ETH               & RDN             & BNT                & GNO              \\
\textbf{From:}     & \textbf{To:}    & RDN               & ETH             & GNO                & BNT              \\ \cline{1-2}
ETH                & AMN             & ETH               & RLC             & BNT                & HEDG             \\
AMN                & ETH             & RLC               & ETH             & HEDG               & BNT              \\
ETH                & AMPL            & ETH               & SAI             & BNT                & KNC              \\
AMPL               & ETH             & SAI               & ETH             & KNC                & BNT              \\
ETH                & ANT             & ETH               & SAN             & BNT                & MANA             \\
ANT                & ETH             & SAN               & ETH             & MANA               & BNT              \\
ETH                & BAT             & ETH               & SNT             & BNT                & MKR              \\
BAT                & ETH             & SNT               & ETH             & MKR                & BNT              \\
ETH                & BNT             & ETH               & TKN             & BNT                & POA20            \\
BNT                & ETH             & TKN               & ETH             & POA20              & BNT              \\
ETH                & DAI             & ETH               & TRST            & BNT                & RCN              \\
DAI                & ETH             & TRST              & ETH             & RCN                & BNT              \\
ETH                & DATA            & ETH               & UBT             & BNT                & RDN              \\
DATA               & ETH             & UBT               & ETH             & RDN                & BNT              \\ \cline{3-4}
ETH                & ENJ             & \multicolumn{2}{c}{\textbf{Bancor}} & BNT                & RLC              \\
ENJ                & ETH             & \textbf{From:}    & \textbf{To:}    & RLC                & BNT              \\ \cline{3-4}
ETH                & FXC             & BNT               & AMN             & BNT                & SAI              \\
FXC                & ETH             & AMN               & BNT             & SAI                & BNT              \\
ETH                & GNO             & BNT               & AMPL            & BNT                & SAN              \\
GNO                & ETH             & AMPL              & BNT             & SAN                & BNT              \\
ETH                & HEDG            & BNT               & ANT             & BNT                & SNT              \\
HEDG               & ETH             & ANT               & BNT             & SNT                & BNT              \\
ETH                & KNC             & BNT               & BAT             & BNT                & TKN              \\
KNC                & ETH             & BAT               & BNT             & TKN                & BNT              \\
ETH                & MANA            & BNT               & DATA            & BNT                & TRST             \\
MANA               & ETH             & DATA              & BNT             & TRST               & BNT              \\
ETH                & MKR             & BNT               & ENJ             & BNT                & UBT              \\
MKR                & ETH             & ENJ               & BNT             & UBT                & BNT              \\ \cline{5-6} 
ETH                & POA20           & BNT               & ETH             & \multicolumn{2}{c}{\textbf{MakerDAO}} \\
POA20              & ETH             & ETH               & BNT             & \textbf{From:}     & \textbf{To:}     \\ \cline{5-6} 
ETH                & RCN             & BNT               & FXC             & DAI                & SAI              \\
RCN                & ETH             & FXC               & BNT             & SAI                & DAI              \\
\bottomrule
\end{tabular}
\caption{List of the supported DeFi actions of \tool.}
\label{tab:actions}
\end{table}

\section{SMT encoding example}\label{app:encoding_example}
To ease the understanding of the encoding process between the State Transition Model and the SMT problem, we consider in the following a simple strategy with only two actions, and a trader holding two cryptocurrency assets: a \emph{base} cryptocurrency asset $c_1$, and another cryptocurrency asset $c_2$.
\point{Action $a_1$} Converts $x_1$ amount of $c_1$ to $c_2$, using a constant product market (cf.\ Section~\ref{sec:defi-platform-background}), with liquidity $L1^{c_1}$ for $c_1$ and $L1^{c_2}$ for $c_2$ (cf.\ Equation~\ref{eq:a1}).
\begin{equation}\label{eq:a1}
    \text{output amount of } c_2 = L1^{c_2} - \frac{L1^{c_1} L1^{c_2}}{(L1^{c_1} + x_1)}
\end{equation}
\point{Action $a_2$} Converts $x_2$ amount of $c_2$ back to $c_1$, using another constant product market, with liquidity $L2^{c_1}$ and $L2^{c_2}$. Based on Heuristic 5 (cf. Section~\ref{sec:pruning}), action $a_2$ must use another market, because otherwise the conversion becomes a reversing action of $a_1$, which would result in a zero-sum game with a loss on transaction fees.
\point{Initial state encoding}
Equation~\ref{eq:encoding1} encodes the state variables with concrete values, which are fetched from the considered blockchain state (e.g., the most recent block). This predicate can also be viewed as the assignment of an initial state.

\begin{equation}\label{eq:encoding1}
\begin{aligned}
\text{predicate} & \ t_1(\cdot) := \\
& \mathcal{B}^\mathbb{T}_0(c_1) = \text{Trader's initial } c_1 \text{ balance} \ \land \\
& \mathcal{B}^\mathbb{T}_0(c_2) = \text{Trader's initial } c_2 \text{ balance} \ \land \\
& L1_0^{c_1} = \text{Market 1 initial } c_1 \text{ balance} \ \land \\
& L1_0^{c_2} = \text{Market 1 initial } c_2 \text{ balance} \ \land \\
& L2_0^{c_1} = \text{Market 2 initial } c_1 \text{ balance} \ \land \\
& L2_0^{c_2} = \text{Market 2 initial } c_2 \text{ balance} \\
\end{aligned}
\end{equation}

\point{Action encoding}
The following two predicates encode the two state transition actions. Equation~\ref{eq:encoding2} encodes $\mathcal{F}(s_0, a_1, x_1)$ and Equation~\ref{eq:encoding3} encodes $\mathcal{F}(\mathcal{F}(s_0, a_1, x_1), a_2, x_2)$. Simply speaking, predicate $t_2$ transacts cryptocurrency asset $c_1$ to $c_2$, and predicate $t_3$ converts $c_2$ back to $c_1$.

\begin{equation}\label{eq:encoding2}
\begin{aligned}
\text{predicate} & \ t_2(\cdot) := \\
& 0 \leq x_1 \leq \mathcal{B}_0^\mathbb{T}(c_1) \ \land \\
& \mathcal{B}^\mathbb{T}_1(c_1) = \mathcal{B}^\mathbb{T}_0(c_1) - x_1 \ \land \\
& \mathcal{B}^\mathbb{T}_1(c_2) = \mathcal{B}^\mathbb{T}_0(c_2) + L1_0^{c_2} - \frac{L1_0^{c_1} L1_0^{c_2}}{(L1_0^{c_1} + x_1)} \ \land \\
& L1_1^{c_1} = L1_0^{c_1} + x_1 \ \land \\
& L1_1^{c_2} = \frac{L1_0^{c_1} L1_0^{c_2}}{(L1_0^{c_1} + x_1)} \ \land \\
& L2_1^{c_1} = L2_0^{c_1} \ \land \\
& L2_1^{c_2} = L2_0^{c_2} \\
\end{aligned}
\end{equation}

\begin{equation}\label{eq:encoding3}
\begin{aligned}
\text{predicate} & \ t_3(\cdot) := \\
& 0 \leq x_2 \leq \mathcal{B}_1^\mathbb{T}(c_2) \ \land \\
& \mathcal{B}^\mathbb{T}_2(c_1) = \mathcal{B}^\mathbb{T}_1(c_1) + L2_1^{c_1} - \frac{L2_1^{c_1} L2_1^{c_2}}{(L2_1^{c_2} + x_2)}\ \land \\
& \mathcal{B}^\mathbb{T}_2(c_2) = \mathcal{B}^\mathbb{T}_1(c_2) - x_2 \ \land \\
& L1_2^{c_1} = L1_1^{c_1} \ \land \\
& L1_2^{c_2} = L1_1^{c_2} \ \land \\
& L2_2^{c_1} = \frac{L2_1^{c_1} L2_1^{c_2}}{(L2_1^{c_2} + x_2)} \ \land \\
& L2_2^{c_2} = L2_1^{c_2} + x_2 \\
\end{aligned}
\end{equation}

\point{Objective encoding}

We use $Z$ to denote the targeted adversarial revenue. Equation~\ref{eq:encoding4} encodes the objective constraints, ensuring that the adversarial cryptocurrency asset portfolio increases in value. Note that we rely on search algorithms (cf. Algorithm~\ref{alg:dependency}) to find the highest possible $Z$. The optimization process requires solving the same SMT problem with different concrete initialization of revenue targets $Z$ (predicate $t_4$). 

\begin{equation}\label{eq:encoding4}
\begin{aligned}
\text{predicate} & \ t_4(\cdot) := \\
& \mathcal{B}^\mathbb{T}_0(c_1) >= \mathcal{B}^\mathbb{T}_2(c_1) + Z \ \land \\
& \mathcal{B}^\mathbb{T}_0(c_2) = \mathcal{B}^\mathbb{T}_2(c_2)\\
\end{aligned}
\end{equation}

\point{Free variables and range}
Our model only consists of two free variables ($x_1, x_2$) for the simple two action paths. For a path of arbitrary length $n$, the corresponding SMT system consists of $n$ free variables, which are the parameters of each action. As shown in predicate $t_2$ (cf. Equation~\ref{eq:encoding2}) and $t_3$ (cf. Equation~\ref{eq:encoding3}), the range of free variables are constraint by the amount of $\mathbb{T}$'s cryptocurrency assets.

\point{SMT problem}
By following the above procedures, the state transition model we presented in Section~\ref{sec:defi_modeling} is now encoded as an SMT problem, where we verify if any initialization of the free variables ($x_1, x_2$) satisfies the requirement of $t_1(\cdot) \land t_2(\cdot) \land t_3(\cdot) \land t_4(\cdot)$.

\section{Z3 path pruning}

\begin{table}[tb]
\centering
\begin{tabular}
{
>{\raggedleft\arraybackslash}p{2cm}%
>{\raggedleft\arraybackslash}p{2cm}%
>{\raggedleft\arraybackslash}p{2cm}%
}
\toprule
\bf Number of paths SMT must solve & \bf Number of blocks & \bf Percentage of blocks \\
\midrule
0-23        & 0          & 0\%   \\
24          & 204,901    & 21.57\% \\
46          & 609        & 0.06\% \\
47          & 12,201     & 1.28\% \\
48          & 57,265     & 6.03\% \\
50-100      & 35,771     & 3.77\% \\
>100        & 3,897      & 0.41\% \\
total       & 314,644    & 33.12\% \\
\bottomrule
\end{tabular}
\caption{After we apply the dependency-based blockchain state reduction we show in this Table the number of paths the SMT solver must solve. \StateChangeBelowOneHundredPathsBlockPercentage of the blockchain blocks between \StartingBlock and \EndingBlock have less than $100$ ``state changing'' paths, allowing to reduce the SMT computation.}
\label{tab:paths-per-block}
\end{table}

Table~\ref{tab:contract_dependency} illustrates the state change frequency of the top $15$ most frequently changed DeFi markets we consider in this work. The Uniswap DAI market is significantly more active than the other markets, with a state change frequency of~$27.01\%$ of the blocks, while the majority~($78.72\%$) of markets experience a frequency below $2\%$ of the blockchain blocks. Note that every market is only involved in a subset of the~\NumPaths kept strategies after pruning. For example, only~$48$ out of the~\NumPaths strategies involve the Uniswap DAI market.

\begin{table}[tb]
\centering
\setlength{\tabcolsep}{2pt}
\begin{tabular}{p{2.9cm}rr}
\toprule
 \bf Contract &  \bf Count & \bf State change frequency  \\ \midrule
  Uniswap DAI &  28,464 &                               27.01\% \\
   Bancor ETH &  16,466 &                               15.63\% \\
  Uniswap UBT &  13,623 &                               12.93\% \\
  Uniswap MKR &   5,984 &                                5.68\% \\
  Uniswap SAI &   5,195 &                                4.93\% \\
  Uniswap BAT &   5,090 &                                4.83\% \\
  Uniswap KNC &   4,141 &                                3.93\% \\
 Uniswap DATA &   3,546 &                                3.36\% \\
  Bancor DATA &   2,309 &                                2.19\% \\
  Uniswap SNT &   2,300 &                                2.18\% \\
  Uniswap ANT &   1,759 &                                1.67\% \\
   Bancor UBT &   1,714 &                                1.63\% \\
   Bancor ENJ &   1,602 &                                1.52\% \\
  Uniswap ENJ &   1,337 &                                1.27\% \\
 Uniswap MANA &   1,129 &                                1.07\% \\
  Uniswap RLC &   1,073 &                                1.02\% \\
        Other &   9,650 &                                9.16\% \\ \bottomrule
\end{tabular}
\caption{
State pruning statistics, showing that the Uniswap DAI contract experiences the highest state change frequency ($27.01\%$ of blocks).
}
\label{tab:contract_dependency}
\end{table}

\section{Concrete encoding example for Z3}\label{app:concrete}
In this section, we provide a running example to demonstrate the encoding process of \toolZThree. The example performs an arbitrage at block $9,680,000$, which first converts ETH to BNT on Bancor and then converts BNT back to ETH on Uniswap.

\subsection{Initial state encoding}
The initial state encoding consists of the predicates for both the trader $\mathbb{T}$'s initial balances, as well as the initial states of the underlying platforms.

\begin{scriptsize}
\begin{lcverbatim}
# Trader's initial state.
# We assume the trader holds 1000 ETH at the start.
S0_Attacker[BNT] == 0,
S0_Attacker[ETH] == 1000000000000000000000,
\end{lcverbatim}
\end{scriptsize}

\begin{scriptsize}
\begin{lcverbatim}
# Initial states of the underlying platforms.
S0_Uniswap[BNT]_eth == 135368255883939133529,
S0_Uniswap[BNT]_erc20 == 108143877658121296155075,
S0_Bancor[ETH]_erc20 == 10936591981278719837125,
S0_Bancor[ETH]_erc20_ratio == 500000,
S0_Bancor[ETH]_bnt == 8792249012668956788248921,
S0_Bancor[ETH]_bnt_ratio == 500000,
S0_Bancor[ETH]_fee == 1000,
\end{lcverbatim}
\end{scriptsize}

\subsection{Action encoding}

We encode the two transition actions as predicates. $P1$ is the input parameter for action $1$ (converts ETH to BNT on Bancor), and $P2$ is the input parameter for action $2$ (converts BNT to ETH on Uniswap).

\begin{scriptsize}
\begin{lcverbatim}
# converts ETH to BNT on Bancor
P1 > 0,
S1_Bancor[ETH]_bnt > 0,
S1_Attacker[BNT] ==
    S0_Attacker[BNT] +
    (S0_Bancor[ETH]_bnt*
    (1 -
    (S0_Bancor[ETH]_erc20/(S0_Bancor[ETH]_erc20 + P1))**
    (S0_Bancor[ETH]_erc20_ratio/S0_Bancor[ETH]_bnt_ratio))*
    (1000000 - S0_Bancor[ETH]_fee)**2)/
    1000000000000,
S1_Attacker[ETH] == S0_Attacker[ETH] - P1,
S1_Uniswap[BNT]_eth == S0_Uniswap[BNT]_eth,
S1_Uniswap[BNT]_erc20 == S0_Uniswap[BNT]_erc20,
S1_Bancor[ETH]_bnt ==
    S0_Bancor[ETH]_bnt -
    (S0_Bancor[ETH]_bnt*
    (1 -
    (S0_Bancor[ETH]_erc20/(S0_Bancor[ETH]_erc20 + P1))**
    (S0_Bancor[ETH]_erc20_ratio/S0_Bancor[ETH]_bnt_ratio))*
    (1000000 - S0_Bancor[ETH]_fee)**2)/
    1000000000000,
S1_Bancor[ETH]_bnt_ratio == S0_Bancor[ETH]_bnt_ratio,
S1_Bancor[ETH]_erc20_ratio == S0_Bancor[ETH]_erc20_ratio,
S1_Bancor[ETH]_erc20 == S0_Bancor[ETH]_erc20 + P1,
S1_Bancor[ETH]_fee == S0_Bancor[ETH]_fee,
\end{lcverbatim}
\end{scriptsize}

\begin{scriptsize}
\begin{lcverbatim}
# converts BNT to ETH on Uniswap
S1_Attacker[BNT] >= P2,
P2 > 0,
S2_Attacker[BNT] == S1_Attacker[BNT] - P2,
S2_Attacker[ETH] ==
    S1_Attacker[ETH] +
    (997*P2*S1_Uniswap[BNT]_eth)/
    (S1_Uniswap[BNT]_erc20*1000 + 997*P2),
S2_Uniswap[BNT]_eth ==
    S1_Uniswap[BNT]_eth -
    (997*P2*S1_Uniswap[BNT]_eth)/
    (S1_Uniswap[BNT]_erc20*1000 + 997*P2),
S2_Uniswap[BNT]_erc20 == S1_Uniswap[BNT]_erc20 + P2,
S2_Bancor[ETH]_bnt == S1_Bancor[ETH]_bnt,
S2_Bancor[ETH]_bnt_ratio == S1_Bancor[ETH]_bnt_ratio,
S2_Bancor[ETH]_erc20_ratio == S1_Bancor[ETH]_erc20_ratio,
S2_Bancor[ETH]_erc20 == S1_Bancor[ETH]_erc20,
S2_Bancor[ETH]_fee == S1_Bancor[ETH]_fee,
\end{lcverbatim}
\end{scriptsize}

\subsection{Objective encoding}

In this example, we check if it is possible for the trader $\mathbb{T}$ to realize $1$ ETH of revenue following this path.

\begin{scriptsize}
\begin{lcverbatim}
# Objective encoding
S2_Attacker[BNT] == 0,
S2_Attacker[ETH] >= 1001000000000000000000
\end{lcverbatim}
\end{scriptsize}

\section{Optimizer for the SMT solver}

Algorithm~\ref{alg:optimiser} shows how the SMT solver can maximize a path's revenue using binary search.

\begin{algorithm}[]
\SetAlgoLined
\DontPrintSemicolon
\SetKwProg{Fn}{Function}{ is}{end}
\KwIn{\;
\textit{p} $\gets$ Path\;
\textit{m} $\gets$ Minimum revenue target\;
}
\KwOut{Optimized revenue \textit{r}}
\eIf{$\lnot$ \text{isSAT}(\textit{p}, \textit{m})}{
    \Return{0}\;
} {
    $\text{l} \gets m $\;
    $\text{u} \gets m \times 10$\;
}

\While{\text{isSAT}(\textit{p}, \textit{u})}{
    $\text{l} \gets \text{u}$\;
    $\text{u} \gets \text{u} \times 10$\;
}

\Return{binarySearch(\textit{p}, \textit{l}, \textit{u})}\;
\;
\Fn{isSAT(\textit{p}, \textit{r}) : bool}{return \textit{(Is the path \textit{p} SAT for the revenue \textit{r})} \;}
\;
\Fn{binarySearch(\textit{p}, \textit{l}, \textit{u}) : float}{
    Binary search SAT solution on path \textit{p}, using the lower bound \textit{l} and upper bound \textit{u} on revenue \;
    return \textit{(Maximum SAT revenue)} \;
}
\caption{Maximize a path's revenue using SMT solver and binary search.}
\label{alg:optimiser}
\end{algorithm}

\section{State Dependency}
\label{ref:state_dependency}

\begin{figure*}[tb]
\centering
\includegraphics[width=\textwidth]{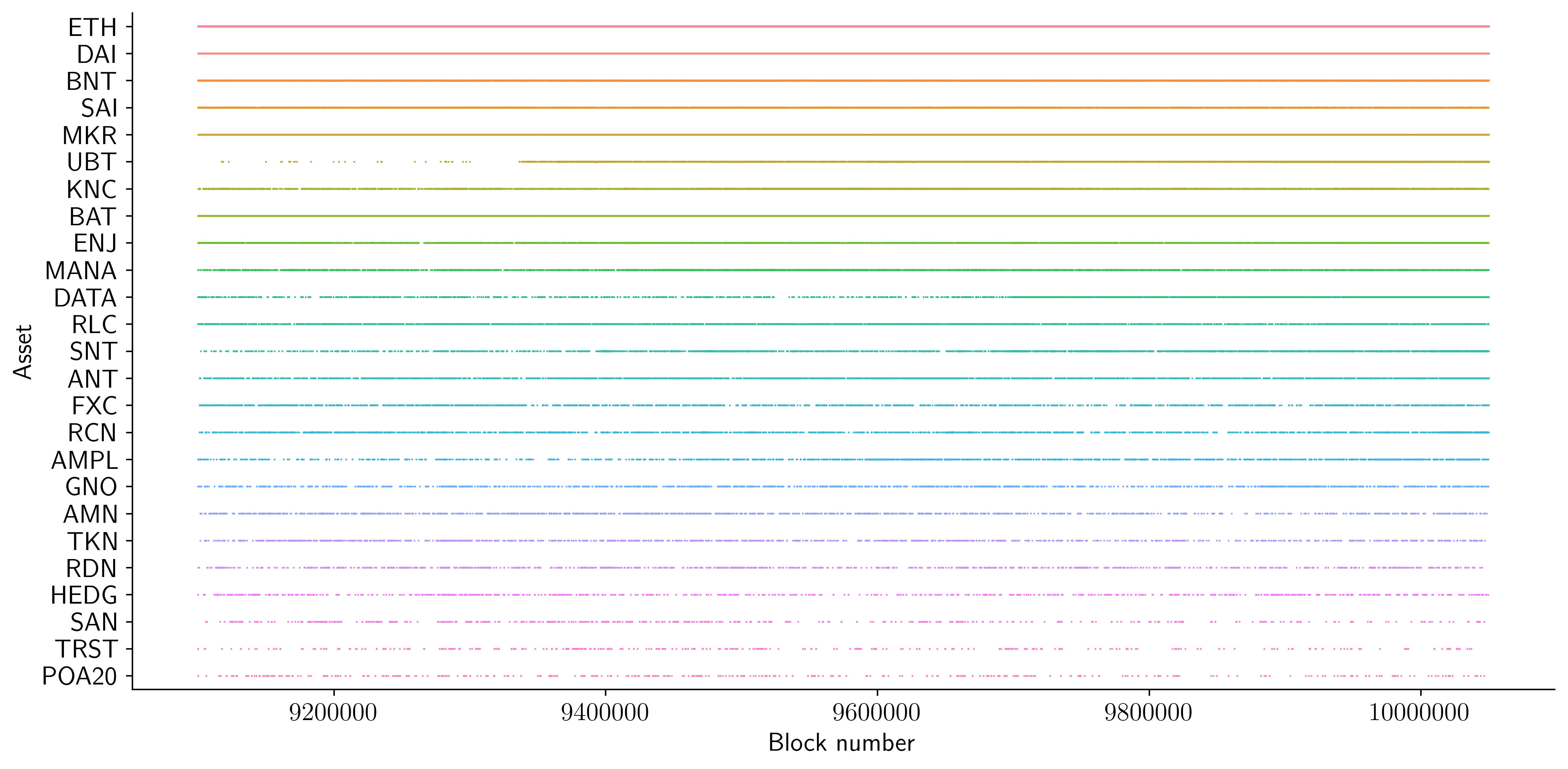}
\caption{
Timeline analysis of the state changes, over~\NumDays~(\NumBlocks blocks), where every state change is represented with a colored tick.}
\label{fig:asset_dependency}
\end{figure*}

We visualize the state changes in Figure~\ref{fig:asset_dependency}. This figure provides an intuition to a trader on how active a particular market is. An asset changes state if a market listing that asset changes state (i.e., a trader trades the asset). ETH experiences the most state changes with over~\NumBlocks blocks~(\ETHStateChange). After ETH, we observe that DAI~(\DAIStateChange) experiences the most frequent state changes over the~\NumBlocks blocks we crawled. POA20 has the lowest number of state changes~(\POAStateChange). For a trader who is not able to position its transactions first in a block, the market activity is relevant because a strategy executed on the POA20 asset has a higher likelihood to succeed than on an active DAI market. 

\section{bZx}

\begin{figure*}[htb!]
\centering
\includegraphics[width = \textwidth]{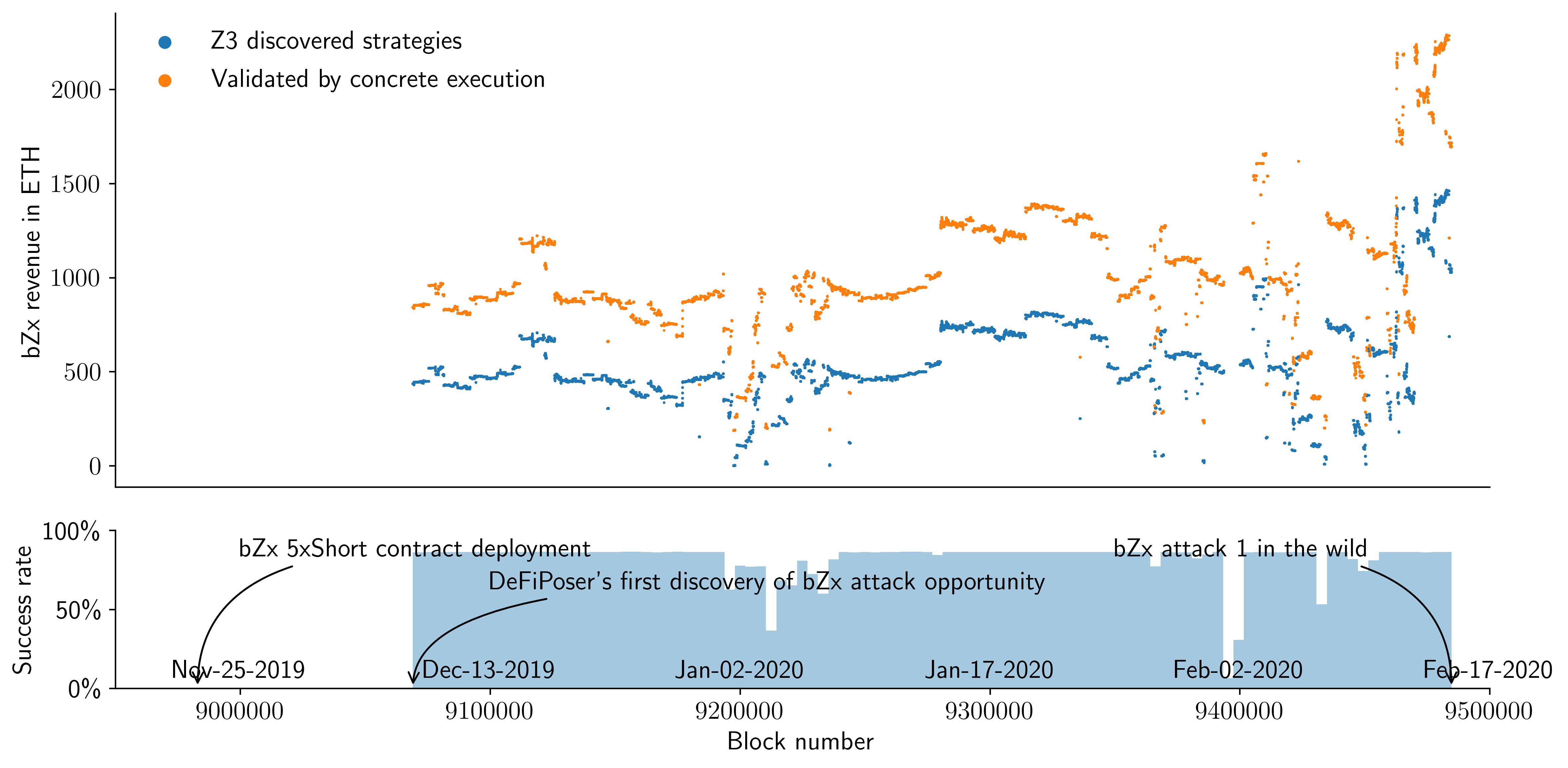}
\caption{Attack window analysis of the bZx attack. \toolZThree finds the first attack opportunity at block~$9,069,000$ (December 8th 2019). The opportunity lasted for~$69$ days, until the opportunity was exploited in block~$9,484,687$ (February 15th 2020). We visualize the difference between the profits from Z3 and concrete validation, along with the success rate (using block bin sizes of $100$) of a Z3 strategy passing concrete validation. Note that the bZx loan interest rate formula is conservatively simplified in the encoding process, which explains why the Z3 anticipated revenue is lower than the concrete execution yield.}
\label{fig:bzx_attack_window}
\end{figure*}

Figure~\ref{fig:bzx_attack_window} shows our attack window analysis of the bZx attack using \toolZThree.

\end{document}